%% file: main.tex
\documentclass[12pt, a4paper, french]{report}
\usepackage[utf8]{inputenc}
\usepackage[T1]{fontenc}
\usepackage[francais, english]{babel}
\usepackage{mathtools,amssymb,amsthm}
\usepackage{lmodern}
\usepackage{ragged2e}
\usepackage{geometry}
\usepackage{color}
\usepackage{float}
\usepackage{bold-extra}

\usepackage{pdfpages}
\usepackage{subcaption}
 
\usepackage{nomencl}
\makenomenclature

\usepackage{diagbox}
\usepackage{lscape}
\newlength\algowd

\let\oldnl\nl
\newcommand{\nonl}{\renewcommand{\nl}{\let\nl\oldnl}}

\usepackage{graphicx}
\usepackage[table, dvipsnames, x11names, svgnames]{xcolor}
\usepackage{array,multirow,makecell}
\setcellgapes{1pt}
\makegapedcells
\newcolumntype{R}[1]{>{\raggedleft\arraybackslash }b{#1}}
\newcolumntype{L}[1]{>{\raggedright\arraybackslash }b{#1}}
\newcolumntype{C}[1]{>{\centering\arraybackslash }b{#1}}
\usepackage{calc}
\usepackage[table]{xcolor}
\usepackage{url}
\usepackage{titlesec}
\usepackage{fancyhdr}
\usepackage{enumitem}

\usepackage[export]{adjustbox}
\usepackage{wrapfig}
\usepackage[justification=centering]{subcaption}

\usepackage{multicol}
\usepackage{indentfirst}
\usepackage[autolanguage]{numprint}
\usepackage{placeins}
\let\Oldsection\section
\renewcommand{\section}{\FloatBarrier\Oldsection}
\let\Oldsubsection\subsection
\renewcommand{\subsection}{\FloatBarrier\Oldsubsection}
\let\Oldsubsubsection\subsubsection
\renewcommand{\subsubsection}{\FloatBarrier\Oldsubsubsection}
\usepackage{multirow}

\usepackage{amsmath}
\usepackage[ruled,vlined,french,titlenumbered]{algorithm2e}

\usepackage{setspace}
\titleformat{\chapter}
[frame]
{\normalfont}
{\filright\large\itshape\enspace
Chapitre \thechapter\enspace}
{10pt}
{\Large\bfseries\filcenter}
\usepackage{booktabs}
\usepackage{multirow}
\usepackage{siunitx}
\usepackage{array} 
\usepackage[french]{minitoc}
\dominitoc

\begin{document}
\begin{titlepage}
\newcommand{\HRule}{\rule{\linewidth}{0.5mm}} 

\center
\textbf{UNIVERSITÉ DE TUNIS EL MANAR} \\
\vspace{0.1em}
\textbf{FACULTÉ DES SCIENCES MATHÉMATIQUES, PHYSIQUES ET NATURELLES DE TUNIS}
\vspace{1em}
\begin{figure}[h!]
\centering
\includegraphics[width=0.3\textwidth]{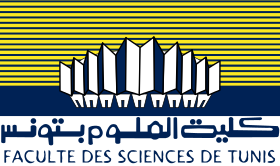}

\end{figure}
\vspace{2.3em} \\
{\bfseries \Large Mémoire de Mastère } \\
\vspace{0.1em}
{\bfseries \normalsize Présenté en vue de l'obtention du diplôme de } \\
\vspace{1em}
{\textrm {\large \textbf{MASTÈRE DE RECHERCHE EN INFORMATIQUE}}} \medskip

{\bfseries \normalsize Par } \medskip

\vspace{0.1em}
{\bfseries \large 	Meriam MENAI} \\
\vspace{3em}

\HRule\\[0.9cm]
	
	{\textrm { \large \textbf{UNE APPROCHE DE D\'ETECTION DE SUJETS BAS\'EE SUR LES R\'ESEAUX DE NEURONES ADAPTATIFS.}}}\\[0.4cm]

\HRule\\[0.5cm]

\vspace{0.3em}

\begin{table}[H]
\resizebox{\linewidth}{!}{
\begin{tabular}{lll}
\multicolumn{3}{l}{Soutenu le 22/12/2018, devant le jury composé de:}   \\ \\
\textbf{M. Sadok BEN YAHIA}   & \textbf{Professeur, FST}          & \textbf{\textit{Président du jury}}     \\
\textbf{M. Sami ZGHAL}        & \textbf{Maître Assistant, FSJEGJ} & \textbf{\textit{Rapporteur}           } \\
\textbf{Mme Chiraz TRABELSI} & \textbf{Maître Assistante, ISAMM} & \textbf{\textit{Directrice du mémoire}}
\end{tabular}}
\end{table}

\bigskip

\bigskip

\bigskip

\bigskip

\bigskip

\bigskip

\bigskip

\bigskip

\textbf{ Année universitaire : 2017-2018 }

\pagenumbering{gobble}
\end{titlepage}
\newpage
\vspace*{\stretch{1}}
\begin{center}
\noindent {\Large{\textit{\textbf{REMERCIEMENTS}}}}
\end{center}

Avant tout développement sur cette expérience, il apparaît opportun de commencer ce rapport par des remerciements, à ceux qui m’ont beaucoup appris au cours de ce mastère, et même à ceux qui ont eu la gentillesse d’en faire un moment très profitable.\bigskip

Mes remerciements vont également à \textbf{Madame Chiraz TRABELSI}, Maître Assistante à l'Institut Supérieur des Arts Multimédia de la Manouba, ma directrice de mémoire, son encadrement, ses conseils et encouragements qu’elle m’a afflué.\bigskip

Je tiens à exprimer ma profonde gratitude à \textbf{Monsieur Sadok BEN YAHIA}, Professeur à
la Faculté des Sciences de Tunis, pour l’honneur qu’il m’a fait en acceptant de présider le jury de soutenance.\bigskip

Mes remerciements s’adressent aussi à \textbf{Monsieur Sami ZGHAL}, Maître Assistant à la Faculté des Sciences Juridiques, Économiques et de Gestion de Jendouba, d’avoir accepté de rapporter mon travail avec patience et pertinence.\bigskip

Mes derniers remerciements et non les moindres, iront à mes parents qui m’ont toujours apporté leur soutien sans failles et à tous ceux qui m’ont soutenue de près ou de loin à réaliser ce travail.

\begin{flushright}
\noindent \textit{\textbf{Merci.}}
\end{flushright}
\vspace*{\stretch{1}}
\vspace*{\stretch{1}}

\selectlanguage{francais}

\newpage

\noindent
\begin{spacing}{1.5} \tableofcontents
\end{spacing}
\newpage

\noindent
\begin{spacing}{1.5} \listoffigures
\end{spacing}
\newpage

\noindent
\begin{spacing}{1.5} \listoftables
\end{spacing}
\newpage
\printnomenclature 

\pagenumbering{arabic}
\addcontentsline{toc}{chapter}{Introduction générale}
\chapter*{Introduction générale}
\minitoc
Avec la quantité croissante de données dans plusieurs domaines, les utilisateurs ont accès à une multitude de données. La récupération d'informations à partir de ces données est un défi lorsque celles-ci sont semi-structurées. Ces données sont parfois sous forme d'articles de nouvelles. Ces derniers sont une source d'information commune, qui peut être trouvée sur différentes sources d'information telles que les journaux en ligne, les blogs ou d'autres types de sites d'information. Cependant, plusieurs sources d'information peuvent couvrir la même catégorie de nouvelles et par conséquent publier des articles similaires, c'est-à-dire qu'elles couvrent les mêmes informations. Il serait souhaitable que le lecteur puisse accéder facilement à tous les articles similaires.\bigskip

En effet, pour qu'un lecteur accède aux articles similaires, il serait souhaitable qu'ils soient regroupés ensemble dans plusieurs catégories. Les méthodes classiques de catégorisation exploitent l’information contenue dans le document pour déterminer sa classe. Dans la majorité des cas, ce sont les mots qui sont utilisés pour représenter cette information. Le problème du clustering de textes a été largement étudié. Le clustering de textes est la tâche qui consiste à assigner un ou plusieurs groupes à un document $d_{i}$. Pour cela, nous disposons d’un corpus dit "d’apprentissage" composé d’un ensemble de documents dont le(s) groupes(s) d’appartenance sont inconnus. Le système de détection de sujets est ensuite entraîné sur cet ensemble d’apprentissage dans le but de correctement catégoriser un nouveau document. Dans notre cas, une classe peut être assimilée à un sujet, l’objectif étant de retrouver le sujet $c_{k}$ du document $d_{i}$.\bigskip

D’autres approches exploitent les réseaux de neurones pour déterminer l’appartenance des documents. En effet, dans les approches par réseaux de neurones, le document à classer est présenté à l’entrée du réseau. La couche de sortie, quant à elle, représente l’ensemble des classes. Après activation du réseau, les valeurs de la couche de sortie représentent les classes possibles du document. Cependant, à l’ère des données massives, la surcharge d’information constitue un problème urgent à considérer. Les systèmes d’apprentissage font, en effet, face au dilemme de plasticité-stabilité. Un système d’apprentissage doit être adaptatif pour réagir aux changements de l’environnement et doit être en même temps stable pour préserver les connaissances acquises précédemment.\bigskip

Les modèles proposés pour le clustering des documents basé sur les réseaux de neurones adaptatifs sont encore à un stade précoce, ils présentent ainsi certaines limites. Ils nécessitent une quantité très importante de données d’apprentissage afin de détecter une bonne représentation qui établit des alignements appropriés entre le document original et le sujet correspondant.\medskip

Nous proposons dans ce mémoire une nouvelle approche de clustering basée sur les réseaux de neurones adaptatifs ART (Adaptive Resonance Theory network). Les réseaux ART sont des réseaux de neurones compétitifs, constitués d’un ensemble de neurones et qui sont à la base d’un modèle d’apprentissage non supervisé. Nous allons exploiter les réseaux de neurones afin de détecter les sujets des documents.

\section*{Structure du mémoire}

Les résultats de nos travaux de recherche sont synthétisés dans ce mémoire qui est composé de quatre chapitres.\medskip

\textbf{\textit{Chapitre 1: Concepts de base}}

Le premier chapitre introduit les notions essentielles pour cerner le champ d’étude.
Nous nous intéressons à la présentation des fondements de la fouille de textes. Nous présentons également les réseaux de neurones et la notion de clustering.\bigskip

\textbf{\textit{Chapitre 2: Revue sur la détection des sujets et la théorie de résonance adaptative}}

Nous présentons dans un premier temps dans ce chapitre, la notion de détection de sujets. Nous introduisons, dans un deuxième temps, quelques approches de détection de sujets existantes dans la littérature. Nous présentons également la théorie de résonance adaptative.\bigskip

\textbf{\textit{Chapitre 3: ClusART: Une approche de détection de sujets basée sur les réseaux de neurones adaptatifs (ART)}}

Le troisième chapitre est consacré à la présentation de notre nouvelle approche de détection de sujets, appelée ClusART. Nous procédons dans une première étape à la présentation de l’architecture générale de ClusART pour décrire par la suite, les différentes phases de ClusART. Dans une deuxième étape, nous présentons le pseudo-algorithme de ClusART.\bigskip

\textbf{\textit{Chapitre 4: Expérimentation et évaluation}}

Durant ce chapitre nous nous consacrons à l’évaluation de notre approche pour la détection de sujets. Nous procédons dans un premier temps, à la présentation de notre environnement d’expérimentations et la description de la collection de test. Dans un deuxième temps, nous proposons une série d’expérimentations visant à promouvoir les performances de notre approche pour la détection de sujets.\bigskip

Le rapport se termine par une conclusion générale qui résume l’ensemble de nos travaux et présente quelques perspectives futures de recherche.
\newtheorem{definition}{Définition}
\chapter{Concepts de base}
\minitoc
\addcontentsline{toc}{section}{Introduction}
\section*{Introduction}
Dans ce chapitre nous procédons tout d’abord par la présentation de la fouille de données. Nous présentons par la suite les notions de base de la fouille de texte. Nous présentons dans une troisième partie la recherche d'information. Nous réservons la dernière partie de ce chapitre à la présentation du soft computing ainsi que quelques unes de ses composantes.
\section{Fouille de données}
\subsection{Définition de la fouille de données} 
La fouille de données (ou exploration de données, en anglais Datamining, DM) réside dans la découverte de connaissances à partir de données (en anglais, knowledge discovery in databases or KDD). C'est un domaine pluridisciplinaire assemblant des techniques d’apprentissage automatique, la reconnaissance des formes, les statistiques, les bases de données et la visualisation.\medskip 

La fouille de données est une activité qui consiste à prélever des informations nouvelles, implicites, non triviales, inconnues au préalable et potentiellement utiles dans des données volumineuses, qui sont généralement sous la forme de régularités: des motifs cachés, des tendances et relations inattendues, etc.\smallskip 

Conformément à la définition de Fayyad \cite{fayyad1996data}, la fouille de données est un processus itératif par lequel on extrait des connaissances valides, nouvelles, potentiellement utiles et compréhensibles en dernière analyse.\smallskip 

Dans une autre définition la fouille de données \cite{hand2007principles} est considérée comme l'analyse de bases de données (La plupart du temps très volumineuses) dans le but de découvrir des relations insoupçonnées et de résumer les données d'une manière à la fois utile et compréhensible.\bigskip

Bien que la fouille de données et la découverte de la connaissance (KDD) soient constamment traitées comme étant synonymes, la fouille de données est véritablement une partie du KDD.\smallskip 

La figure 1.1 suivante illustre le processus d'extraction des connaissances à partir de données et elle montre la fouille de données comme étant une étape dans ce processus.\bigskip

\begin{center}
\begin{figure}[H]
\centering 
\resizebox{\linewidth}{!}{\includegraphics[]{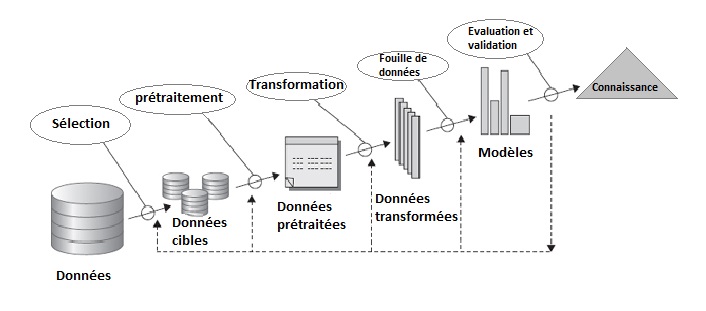}}
\captionof{figure}{Processus de la découverte de la connaissance \cite{fayyad1996data}}
\end{figure}
\end{center}
\subsection{Techniques de la fouille de données}
Les outils de Data Mining emploient des fondements théoriques semblables à ceux des statistiques, de l'apprentissage automatique et de l'intelligence artificielle pour faire face au volume des données continuellement croissant.
Nous présentons cette classification dans le tableau 1.1 qui suit où les méthodes appartenant à la statistique et l'analyse de données traditionnelle apparaissent.\bigskip
\begin{table}[H]
\centering
\caption{Classification des techniques de la fouille de données \cite{salhi2016application}}

\resizebox{\linewidth}{!}{
\begin{tabular}{|l|l|l|l|}
\hline
Type                                                                             & Famille                                                                                                  & Sous-famille                                                                                                & Algorithme                                                                                                                       \\ \hline
\multirow{8}{*}{\begin{tabular}[c]{@{}l@{}}Méthodes\\ descriptives\end{tabular}} & \multirow{5}{*}{\begin{tabular}[c]{@{}l@{}}Modèles\\ géométriques\end{tabular}}                          & \multirow{2}{*}{\begin{tabular}[c]{@{}l@{}}Analyse\\ factorielle\end{tabular}}                              & Analyse en composants principales ACP                                                                                            \\ \cline{4-4} 
                                                                                 &                                                                                                          &                                                                                                             & \begin{tabular}[c]{@{}l@{}}Analyse factorielle des correspondances AFC\\ Analyse des correspondances multiples ACM\end{tabular}  \\ \cline{3-4} 
                                                                                 &                                                                                                          & \multirow{2}{*}{\begin{tabular}[c]{@{}l@{}}Analyse \\ topologique\end{tabular}}                             & \begin{tabular}[c]{@{}l@{}}Méthodes de partitionnement (centres mobiles,\\ K-means, nuées dynamiques, K-medoids...)\end{tabular} \\ \cline{4-4} 
                                                                                 &                                                                                                          &                                                                                                             & Méthodes hiérarchiques                                                                                                           \\ \cline{3-4} 
                                                                                 &                                                                                                          & \begin{tabular}[c]{@{}l@{}}Analyse\\ topologique +\\ réduction de\\ dimension\end{tabular}                  & Classification neuronale(réseaux de Kohonen)                                                                                     \\ \cline{2-4} 
                                                                                 & \begin{tabular}[c]{@{}l@{}}Modèles \\ combinatoires\end{tabular}                                         &                                                                                                             & Classification relationnelle                                                                                                     \\ \cline{2-4} 
                                                                                 & \multirow{2}{*}{\begin{tabular}[c]{@{}l@{}}Modèles de base \\ de règles logiques\end{tabular}}           & \multirow{2}{*}{\begin{tabular}[c]{@{}l@{}}Détection \\ de liens\end{tabular}}                              & Recherche d'associations                                                                                                         \\ \cline{4-4} 
                                                                                 &                                                                                                          &                                                                                                             & Recherche de séquences similaires                                                                                                \\ \hline
\multirow{7}{*}{\begin{tabular}[c]{@{}l@{}}Modèles\\ prédictifs\end{tabular}}    & \begin{tabular}[c]{@{}l@{}}Modèles de \\ base de règles \\ logiques\end{tabular}                         & \begin{tabular}[c]{@{}l@{}}Arbres de\\ décisions\end{tabular}                                               & \begin{tabular}[c]{@{}l@{}}Arbres de décision(variable à expliquer continue \\ ou qualitative)\end{tabular}                      \\ \cline{2-4} 
                                                                                 & \multirow{5}{*}{\begin{tabular}[c]{@{}l@{}}Modèles à\\ base de\\ fonctions\\ mathématiques\end{tabular}} & \begin{tabular}[c]{@{}l@{}}Réseaux de\\ neurones\end{tabular}                                               & Réseaux à apprentissage supervisé                                                                                                \\ \cline{3-4} 
                                                                                 &                                                                                                          & \multirow{4}{*}{\begin{tabular}[c]{@{}l@{}}Modèles\\ paramétriques\\ ou semi \\ paramétriques\end{tabular}} & \begin{tabular}[c]{@{}l@{}}Régression linéaire, modèle linéaire général\\ GLM, régression PLS\end{tabular}                       \\ \cline{4-4} 
                                                                                 &                                                                                                          &                                                                                                             & \begin{tabular}[c]{@{}l@{}}Analyse discriminante de Fisher, régression \\ logistique, régression logistique PLS\end{tabular}     \\ \cline{4-4} 
                                                                                 &                                                                                                          &                                                                                                             & Modèle log-linéaire                                                                                                              \\ \cline{4-4} 
                                                                                 &                                                                                                          &                                                                                                             & \begin{tabular}[c]{@{}l@{}}Modèle linéaire généralisé GLZ, modèle additif\\ généralisé GAM\end{tabular}                          \\ \cline{2-4} 
                                                                                 & \begin{tabular}[c]{@{}l@{}}Prédiction\\ sans modèle\end{tabular}                                         & \begin{tabular}[c]{@{}l@{}}Analyse \\ probabiliste\end{tabular}                                             & K-plus proches voisins(K-NN)                                                                                                     \\ \hline
\end{tabular}}
\end{table}
\medskip

La table 1.1 expose les principaux algorithmes de Data Mining qui sont distribués en deux grandes familles de techniques:\bigskip

\begin{itemize}
\item Les techniques descriptives (pas de variables à expliquer): Leur but est de mettre en évidence des informations qui existent mais qui sont cachées par le volume de données. Elles sont employées pour la réduction, le résumé et la synthèse des données qui ne possèdent pas de variable cible à prédire.
\item Les techniques prédictives (avec une variable à expliquer):  Leur but est d'extrapoler de nouvelles informations à partir des informations présentes pour expliquer les données, elles sont employées aussi pour la prédiction des variables cibles.
\end{itemize}

Pour synthétiser, les techniques prédictives s'intéressent aux variables et aux relations entre elles, tandis que  les techniques descriptives s'intéressent aux individus ainsi qu'aux groupes homogènes qu'ils peuvent former.\bigskip

Il n’existe pas de méthode de fouille de données meilleure que les autres, mais il existe des compromis selon les besoins dégagés et la nature des applications. En conséquence, à tout jeu de données et tout problème correspond une ou plusieurs méthodes. Le choix s’effectue en fonction de la tâche à résoudre, de la nature et de la disponibilité de données, des connaissances, des compétences disponibles et de la finalité du modèle construit. Pour cela, les critères suivants sont importants: la complexité de la construction du modèle, la complexité de son utilisation et ses performances.
\section{Fouille de textes}
\subsection{Définition de la fouille de textes}

Text Mining, fouille de texte ou découverte de connaissances à partir de texte (KDT) a été introduite par Fledman et al. \cite{feldman1995knowledge}, elle est incluse dans le domaine de l'intelligence artificielle. Elle fait référence au processus d'extraction d'informations de haute qualité à partir du texte. La fouille de texte traite les données textuelles non structurées dans un format structuré pour une analyse plus approfondie. On peut la définir comme un processus à forte intensité de connaissances dans lequel un utilisateur interagit avec une collection de documents au fil du temps en utilisant une suite d'outils d'analyse. D'une manière analogue à la fouille de données, la fouille de texte cherche à extraire des informations utiles à partir de sources de données à travers l'identification et l'exploration de modèles intéressants. La fouille de texte peut être résumée en trois étapes, comme indiqué sur la figure 1.2:\bigskip

\begin{itemize}
\item Étape 1: D'abord, une source de données est sélectionnée.
\item Étape 2: Ensuite, ces données sont pré-traitées.
\item Étape 3: Puis, ces données sont analysées. Et à la fin, les résultats sont interprétés.
\end{itemize}
\bigskip

\begin{figure}[H]
\centering 
\resizebox{\linewidth}{!}{\includegraphics[]{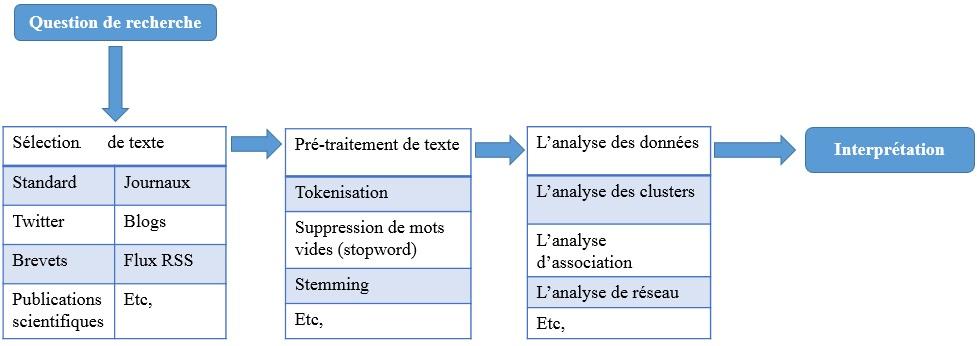}}
\captionof{figure}{Processus de la fouille de textes}
\end{figure}
\subsection{Concepts de la fouille de textes}
La fouille de texte est caractérisée par trois concepts clés \cite{feldman2007text}: \medskip

\begin{itemize}
\item Un corpus ou une collection de documents.
\item Un document unique. 
\item Des caractéristiques documentaires.
\end{itemize}
\bigskip

On peut définir une collection de documents comme un regroupement de documents texte dans lequel nous voulons découvrir des modèles. Pour pouvoir activer la phase de modélisation de la fouille de données, chacun des documents doit être représenté comme un ensemble de caractéristiques de document qu'il contient. Plus généralement, lorsqu'il est question de texte totalement non structuré, les caractéristiques sont procédés du texte en utilisant uniquement des méthodes de prétraitement de texte; toutefois, une caractéristique peut être toute information pouvant être attribuée à un document (par exemple, auteur, date de publication, longueur du document, mots-clés assignés et similaire).\bigskip

Les mots, les termes et les concepts qui apparaissent dans le texte sont les caractéristiques du document les plus couramment utilisées. Les caractéristiques les plus simples sont les mots qui ne portent aucune valeur sémantique supplémentaire et représentent des mots exacts extraits du texte.\bigskip
 
Un pas plus haut sur l'échelle sémantique sont des termes qui sont des ensembles d'un ou plusieurs mots se montrant ensemble et qui sont déjà normalisés, filtrés et regroupés en employant diverses méthodes d'extraction de termes. Les caractéristiques ayant la valeur la plus sémantique sont des concepts qui ne sont pas seulement des termes qui apparaissent dans le texte, mais aussi des notions connexes, mais pas indispensablement mentionnées (par exemple, un document décrivant une voiture spécifique peut ne pas contenir le mot "voiture" la caractéristique de concept "voiture" pourrait être assignée à un tel document).\medskip 

L'une des caractéristiques considérables des caractéristiques du document décrit est le fait qu'elles sont généralement rares \cite{feldman2007text}, ce qui signifie que la plupart des caractéristiques d'un vecteur ont un poids nul. Cette caractéristique de rareté est due au fait qu'il existe de multiples caractéristiques (mots, termes ou concepts) dans la collection de documents; cependant, un seul document ne contient qu'un petit sous-ensemble d'entre eux \cite{jurvsivc2015text}.
\subsection{Historique de la fouille de textes}
La fouille de texte est apparue pendant la deuxième moitié des années 90 aux États-Unis, en écho à des travaux accomplis depuis les années 80 sur des bases de données.
Piatetsky-Shapiro introduit en 1991 comme titre de son ouvrage le terme de Knowledge Discovery from Databases (KDD), en français, Extraction de Connaissances à partir de Bases de Données (ECBD). L’usage des termes Knowledge Discovery from Databases et Data Mining ne se précise que vers l'année 1995.
\section{Recherche d'information}
\subsection{Définition de la recherche d'information}
Avec la progression rapide du volume documentaire conservé sous format numérique, il est devenu très difficile de trouver une information ou un document qui correspond à un besoin utilisateur. La Recherche d’Information (RI) n’est pas une discipline récente, elle date des années 40. Une des premières définitions de la RI a été donnée par Salton: « la recherche d’information est un domaine qui a pour objectif, la représentation, l’analyse, l’organisation, le stockage et l’accès à l’information ».\bigskip

La recherche d’information est un domaine qui a été développé en même temps que les systèmes de base de données depuis de nombreuses années. Il peut être défini comme une activité dont la finalité est de localiser et de délivrer une collection de documents à un utilisateur en fonction de son besoin en informations \cite{bouramoul2011recherche}. Contrairement au domaine des systèmes de base de données, qui s’est focalisé sur le traitement des requêtes et des transactions de données structurées, le défit de la recherche d'information est d'être capable, parmi le volume important de documents disponibles, d'organiser et trouver ceux qui correspondent au mieux à l'espérance de l’utilisateur. En général, un système de Recherche d’Information (RI) prend en entrée une requête énoncée par un utilisateur puis va récupérer des données au sein d’une collection indexée auparavant.\bigskip

Historiquement, la RI fait principalement mention à la recherche documentaire: les
données récupérées sont des documents entiers qui contiennent des informations que le système a jugé comme pertinentes par rapport à la requête \cite{harman2011information}. Le système tente de trouver les documents qui contiennent les mots-clés, afin de procurer à l’utilisateur une liste de documents classés en fonction de leur pertinence estimée vis-à-vis de la requête. Actuellement, la RI \cite{devea2013vers} n’est néanmoins plus limitée à cette recherche documentaire et se rapproche de l’accès à l’information en général. Parmi les multiples aspects de la RI, nous pouvons citer, entre autres, la recherche de passages (uniquement quelques parties des documents) (\cite{kaszkiel1997passage}; \cite{fuhr2007overview}), la génération de mini-phrases décrivant les documents dans la liste de résultats (ou snippets) \cite{huang2008query} ou le résumé multi-documents orienté par une requête \cite{boudin2007neo}.\bigskip

L’informatique a permis de développer des outils pour traiter l’information et établir la représentation des documents au moment de leur indexation, ainsi que pour rechercher l’information. Actuellement on peut dire que la recherche d'information est un champ pluridisciplinaire qui peut être étudié par plusieurs disciplines utilisant des approches qui devraient aider à trouver des solutions pour améliorer son efficacité \cite{bouramoul2011recherche}.\bigskip

L’opérationnalisation de la RI est accomplie par des outils informatiques appelés Systèmes de Recherche d’Information (SRI) \cite{harrathi2010recherche}. Dans un SRI l’utilisateur formule son besoin sous forme de requête et le SRI renvoie l’ensemble des documents qui correspondent à cette requête. Afin de réaliser cela, il met en œuvre deux processus:\bigskip

\begin{itemize}
\item Un processus d’indexation: Le processus d’indexation se résume par la description de chaque document de la collection et chaque requête par un ensemble de descripteurs. Ces descripteurs permettent d'indiquer au mieux le contenu du document ou de la requête. Ainsi, les documents et les requêtes sont représentés dans un même espace dit espace d’indexation. Dans cet espace chaque document et chaque requête dispose d'une représentation interne unique.

\item Un processus d’appariement: Le processus d’appariement ou de comparaison se résume dans la mise en correspondance de la représentation de la requête avec les représentations des documents. Cette mise en correspondance aide à calculer un degré de ressemblance ou de similarité entre chaque requête et chaque document de la collection. En s'appuyant sur ce degré, les documents qui sont jugés similaires (par le SRI) à la requête sont par la suite renvoyés à l’utilisateur. Certains systèmes permettent de présenter ces documents dans une liste triée suivant un ordre décroissant de leur ressemblance avec la requête utilisateur.
\end{itemize}
\bigskip

\begin{figure}[H]
\centering 
\resizebox{12cm}{!}{\includegraphics[]{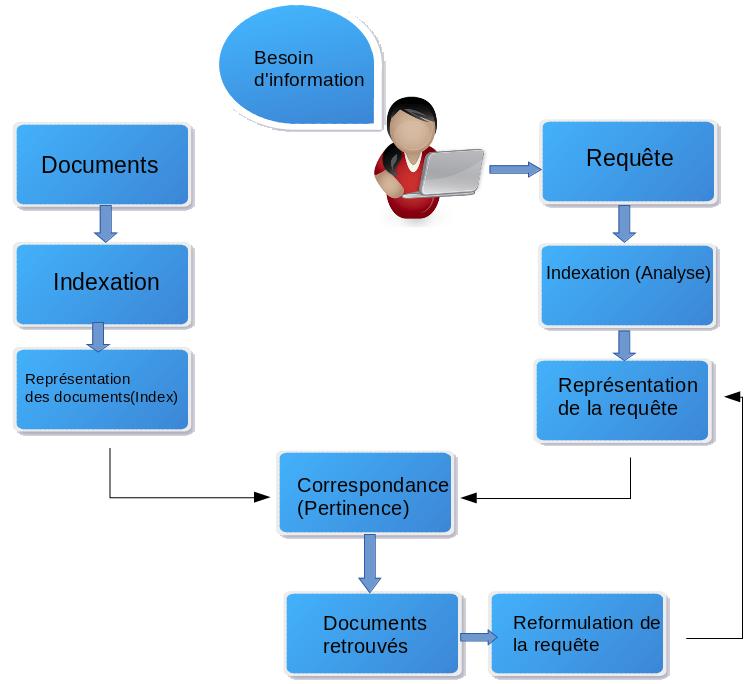}}
\captionof{figure}{Architecture générale d'un système de recherche d'information}
\end{figure}
\subsection{Concepts de base de la recherche d'information}
La recherche d'information s'articule autour d'un certains nombre de concepts clés:
\begin{itemize}
\item \textbf{Une collection de documents:} elle constitue l'ensemble des informations exploitables et accessibles. Elle comprend un ensemble de documents.

\item \textbf{Document:} constitue une granule d'information élémentaire d'une collection de documents. Cette granule de document peut représenter tout ou une partie d'un document.

\item \textbf{Besoin d'information:} la notion de besoin constitue l'ensemble des mots clés par lesquels l'utilisateur décrit sa recherche d'information. Cet ensemble forme une requête qui peut être exprimée sous plusieurs formes.

\item \textbf{Pertinence:} la notion de pertinence représente un critères majeur de l'évaluation des performances d'un système de RI. Elle montre dans quelle mesure les documents retournés par le SRI répondent au besoin d'information de l'utilisateur.

\item \textbf{Requête:} c'est l'expression du besoin en information de l'utilisateur par un ensemble de mots clés. Elle représente l'interface entre le SRI et l'utilisateur. 
\end{itemize}

\section{Calcul souple}
Les dernières années ont connu une croissance rapide de l'intérêt d'un ensemble de modes de modélisation et de calcul qui peuvent être décrits collectivement sous le nom de soft computing. Il a été introduit par L.A. Zadeh en 1994 comme un moyen de construction des systèmes intelligents qui répond à des obligations d'efficacité, de robustesse, de facilité d'implémentation et d'optimisation de coûts temporels, énergétiques, financiers, etc, mais aussi qui prend en compte la composante humaine généralement présente dans les systèmes.\bigskip

La caractéristique distinctive d'un soft computing est que son objectif principal est d'atteindre la traçabilité, la robustesse, le faible coût et le $MIQ$ élevé (quodent de l'intelligence de la machine)(machine intelligence quodent) grâce à une exploitation de la tolérance pour l'imprécision et l'incertitude. Ainsi, dans le soft computing, ce qui est habituellement recherché, c'est une solution approximative à un problème précisément formulé ou plus typiquement, une solution approximative à un problème imprécis. Le défi consiste donc à exploiter la tolérance pour l'imprécision et l'incertitude inhérentes à la plupart des problèmes en concevant des méthodes de calcul qui conduisent à une solution acceptable à faible coût.\bigskip
 
Par sa nature, le soft computing est beaucoup plus proche du raisonnement humain que les modes de calcul traditionnels. À ce stade, les principaux composants du soft computing sont la logique floue (FL), la théorie des réseaux neuronaux (NN) et le calcul évolutionnaire (EC), les méthodes d'optimisation telles que les algorithmes génétiques, la théorie du chaos et certaines parties de la théorie de l'apprentissage. De plus en plus, ces composantes sont utilisées conjointement afin d’exploiter les avantages de chacune tout en compensant ses inconvénients par l'emploi d’une autre dont les propriétés sont complémentaires afin d'acquérir une amélioration évidente des performances et de l'adaptabilité.\bigskip

Parmi les applications importantes, les domaines du soft computing sont les systèmes de contrôle, les systèmes experts, les techniques de compression de données, le traitement d'image et les systèmes de prise de décision.\bigskip
 
Le soft computing peut être employé dans la plupart des grands domaines où la mise au point de systèmes intelligents pose des difficultés, de l’apprentissage à la commande de processus en passant à travers les bases de données ou le traitement d’images. Il existe des applications du soft computing à des problèmes réels dans des domaines aussi variés que la recherche d’information, la fouille de données (data mining), l’aide à la décision, la robotique, le contrôle de systèmes complexes. On peut donc soutenir que c'est le soft computing - plutôt que l'informatique traditionnelle - qui devrait être considéré comme le fondement de l'intelligence artificielle. Dans les années à venir, cela pourrait bien devenir une position largement répandue \cite{zadeh1993fuzzy}.
\subsection{Logique floue}
Dans le monde réel, il existe beaucoup de connaissances floues, c'est-à-dire des connaissances vagues, imprécises, incertaines, ambiguës, inexactes ou probabilistes.
L'homme peut utiliser de telles informations parce que la pensée et le raisonnement humains impliquent fréquemment des informations floues, provenant peut-être de concepts humains intrinsèquement inexacts et d'appariements d'expériences similaires plutôt qu'identiques.
Les systèmes informatiques, basés sur la théorie des ensembles classiques et la logique à deux valeurs, ne peuvent pas répondre à certaines questions, comme le fait l'être humain, parce qu'ils n'ont pas de réponses complètement vraies.\bigskip

Lotfi A. Zadeh, de l'Université de Berkeley en Californie formula en 1965, les bases théoriques de la logique floue [Zadeh, 1965]. Il avait introduit la notion de sous-ensemble flou afin de répondre aux problèmes auxquels font face de nombreux systèmes complexes, qui doivent traiter des informations qui ont une nature imparfaite, son concept de base est de graduer l'appartenance à un ensemble, c'est un bon moyen pour prendre en considération l'inexactitude dans la connaissance et de formaliser le cheminement du raisonnement humain.\bigskip

Lorsqu'on discute de la logique floue, il y a un problème sémantique qui nécessite une clarification. Le terme logique floue est en fait utilisé dans deux sens différents. 
Dans son sens étroit, la logique floue est un système logique qui vise à rendre formel le raisonnement approximatif. En tant que tel, il est enraciné dans une logique à valeurs multiples, mais son ordre du jour est tout à fait différent de celui d'un système logique traditionnel à valeurs multiples. À ce sujet, il convient de noter qu'énormément de concepts  expliquant l'efficacité de la logique floue en tant que logique du raisonnement approximatif n'appartiennent pas aux systèmes logiques traditionnels à valeurs multiples. Parmi ceux-ci on trouve le concept d'une variable linguistique, la forme canonique, la règle floue si-alors, les quantificateurs flous et de tels modes de raisonnement, le raisonnement syllogistique et le raisonnement dispositionnel.\bigskip

Dans un sens plus étendu, la logique floue est presque équivalente à un ensemble flou, l'ensemble flou, comme l'indique son nom, est essentiellement une catégorie de classes aux limites floues. L'ensemble flou est beaucoup plus étendu que la logique floue dans son sens étroit et comprend cette dernière comme l'une de ses branches. Parmi les autres branches de la théorie des ensembles flous se trouvent, par exemple, l'arithmétique floue, la programmation mathématique floue, la topologie floue, la théorie des graphes flous et l'analyse des données floues. Ce qu'il est primordial de reconnaître, c'est que toute théorie claire peut être floue par généralisation de concept d'un ensemble au sein de cette théorie au concept d'ensemble flou. En réalité, il est très envisageable que finalement la plupart des théories seront floues. L'impulsion pour la transition d'une théorie nette à une théorie floue vient du fait que la généralité d'une théorie et son applicabilité aux problèmes du monde réel sont substantiellement améliorées en substituant le concept d'ensemble par celui d'un ensemble flou.\bigskip

Cette notion d'ensemble flou procure un point de départ commode pour la construction d'un cadre conceptuel parallèle à bien des égards au cadre employé dans le cas des ensembles ordinaires, mais plus général que celui-ci et, potentiellement, peut s'avérer avoir un champ d'application beaucoup plus étendu, particulièrement dans les domaines de la classification des modèles et du traitement de l'information.\bigskip

Essentiellement, plutôt que la présence de variables aléatoires, un tel cadre procure une manière naturelle de traiter les problèmes dans lesquels la source d'imprécision est l'absence de critères d'appartenance à une classe nettement définis.\smallskip

Soit $\emph{X}$ un espace de points (objets), avec un élément générique de $\emph{X}$ noté $\emph{x}$. Ainsi $X={x}$.
Un ensemble flou (classe) $\emph{A}$ dans $\emph{X}$ est caractérisé par une fonction d'appartenance (caractéristique) $f_{A}({x})$ qui associe à chaque point de $\emph{X}$ un nombre réel dans l'intervalle $[0, 1]$, avec la valeur de $f_{A}({x})$ à $\emph{x}$ représentant le «grade d'adhésion» de $\emph{x}$ dans $\emph{A}$. Ainsi, plus la valeur de $f_{A}({x})$ est proche de l'unité, plus le degré d'appartenance de $\emph{x}$ dans $\emph{A}$ est élevé. Lorsque $\emph{A}$ est un ensemble au sens ordinaire du terme, sa fonction d'appartenance ne peut prendre que deux valeurs $0$ et $1$, avec $f_{A}({x}) = 1$ ou $0$ selon que $\emph{x}$ appartient ou non à $\emph{A}$. Ainsi, dans ce cas, $f_{A}({x)}$ se réduit à la fonction caractéristique familière d'un ensemble $\emph{A}$. Lorsqu'il est nécessaire de faire la différence entre de tels ensembles et des ensembles flous, les ensembles avec des fonctions caractéristiques à deux valeurs seront appelés ensembles ordinaires ou simplement ensembles.
\subsection{Théorie des réseaux de neurones}
\subsubsection{Définition des réseaux de neurones}
Leur nom est dû aux neurones du cerveau humain qui traitent et transmettent l'information. En effet, c’est la description des processus mentaux faite par les neurobiologistes qui est à l’origine des modèles théoriques des réseaux de neurones artificiels(RNA). Ils sont aussi appelés réseaux connexionnistes ou réseaux neuromimétiques. Ils peuvent être considérés comme un modèle mathématique de traitement réparti, constitué de plusieurs composants de calcul non linéaire (neurones), procédant en parallèle et connectés entre eux par des poids.\medskip
Les neurones artificiels sont souvent utilisés sous forme de réseaux qui se distinguent selon le type de connections entre les neurones, une cinquantaine de types peut être listée. En guise d’exemples nous citons : le perceptron de Rosemblat, les réseaux de Hopfield etc.\bigskip

Ces derniers sont les plus utilisés dans le domaine de la modélisation et de la commande des procédés. Ils sont constitués d’un nombre fini de neurones qui sont disposés sous forme de couches. Les neurones de deux couches adjacentes sont interconnectés par des poids. L’information dans le réseau circule d’une couche à l’autre, on dit qu’ils sont de type «feed-forward». Nous distinguons trois types de couches \cite{ammar2007mise}:\medskip

\begin{itemize}
\item \textbf{Couche d’entrée:} les neurones de cette couche accueillent les valeurs d’entrée du réseau et les font passer aux neurones cachés. Chaque neurone reçoit une valeur, il ne fait pas donc de sommation.

\item \textbf{Couches cachées:} chaque neurone de cette couche accueille l’information de multiples couches antécédentes, effectue la sommation pondérée par les poids, puis la transforme selon sa fonction d’activation qui est en général une fonction sigmoïde. Par la suite, il transmet cette réponse aux neurones de la couche suivante.

\item \textbf{Couche de sortie:} elle joue le même rôle que les couches cachées, la seule différence entre ces deux types de couches est que la sortie des neurones de la couche de sortie n’est liée à aucun autre neurone.
\end{itemize}
\subsubsection{Historique des réseaux de neurones}
Les recherches menées dans le domaine du connexionnisme ont démarré avec l'introduction en 1943 par W.MCCulloch et W.Pitts d’un modèle simplifié de neurone biologique usuellement appelé neurone formel. Ils montrèrent également théoriquement que des réseaux de neurones formels simples soient aptes à réaliser des fonctions logiques, arithmétiques et symboliques complexes.\bigskip

En 1949, D.Hebb initie, dans son ouvrage "The Organization of Behavior", la notion
d'apprentissage. Deux neurones entrant en activité en même temps vont être associés (c'est-à-dire que leurs contacts synaptiques vont être renforcés). On parle de loi de Hebb et d'associationnisme.\bigskip

En 1958, F.Rosenblatt développe le modèle du Perceptron. C'est un réseau de neurones inspiré du système visuel. Il dispose de deux couches de neurones: une couche de perception (sert à recueillir les entrées) et une couche de décision. C’est le premier modèle pour lequel un processus d’apprentissage a pu être défini. S’inspirant du perceptron, Widrow et Hoff, développent, dans la même période, le modèle de l'Adaline (Adaptive Linear Element). Ce dernier sera, par la suite, le modèle de base des réseaux de neurones multi-couches.\bigskip

En 1969, Les recherches sur les réseaux de neurones ont été pratiquement délaissées lorsque M.Minsky et S.Papert ont publié leur livre «Perceptrons» (1969) et prouvé les limites théoriques du perceptron, en particulier, l’impossibilité de traiter les problèmes non linéaires par ce modèle.\bigskip

En 1982, Hopfield développe un modèle qui manipule des réseaux totalement connectés basés sur la règle de Hebb pour définir les notions d'attracteurs et de mémoire associative. 
En 1984 c’est la révélation des cartes de Kohonen avec un algorithme non supervisé basé sur l'auto-organisation et suivi une année plus tard par la machine de Boltzman (1985).
Une révolution survient alors dans le domaine des réseaux de neurones artificiels: une nouvelle génération de réseaux de neurones, capables de traiter avec succès des phénomènes non-linéaires: le perceptron multicouche est dépourvu des défauts mis en évidence par Minsky. Proposé pour la première fois par Werbos, le Perceptron Multi-Couche apparaît en 1986 introduit par Rumelhart, et, simultanément, sous une appellation voisine, chez Le Cun (1985). Ces systèmes reposent sur la rétropropagation du gradient de l’erreur dans des systèmes à plusieurs couches, chacune de type Adaline de Bernard Widrow, proche du Perceptron de Rumelhart\cite{ammar2007mise}. 
\subsection{Algorithmes évolutionnaires}
Les algorithmes évolutionnaires (AE) sont des méta-heuristiques basées sur des métaphores biologiques inspirées des mécanismes d'évolution naturelle darwienne. Selon la théorie darwinienne, les individus les plus aptes survivent à la sélection naturelle et se reproduisent d’une génération à l’autre. En termes d’optimisation, l’évolution se traduit par un processus itératif de recherche de l’optimum dans l’espace de recherche. En effet, pour un problème donné, une solution est un individu et un ensemble de solutions correspond à une population d'individus. Chaque individu peut être appelé chromosome, et chaque chromosome est constitué d'un ensemble de caractéristiques, appelées les gènes. Dans le codage binaire, un gène vaut soit 0 soit 1. L'ensemble des gènes d'un individu est son génotype et l'ensemble du patrimoine génétique d'une espèce est le génome. Les différentes versions d'un même gène sont appelées allèles.\smallskip

Le fonctionnement d'un AE \cite{zidi2006systeme} se base sur les étapes suivantes:
\begin{itemize}
\item  \textbf{Genèse:} c'est la première phase de l'algorithme, dans laquelle la population initiale est construite d'une manière aléatoire ou à travers des résultats issus d'autres techniques d'optimisation.
\item \textbf{Évaluation:} elle consiste à calculer la valeur de la fonction de coût pour chaque individu.
\item \textbf{Sélection:} c'est le choix des éléments les plus adaptés pour la formation de la nouvelle génération.
\item \textbf{Recombinaison et mutation:} c'est une phase de reproduction, dans laquelle une nouvelle population est construite à partir des individus sélectionnés, via des opérateurs de croisement et de mutation.
\item \textbf{Arrêt:} il s'agit d'un test de l'efficacité de l'algorithme, à travers une valeur de la fonction objectif à atteindre, le nombre d'itérations ou le temps d'exécution. La solution courante est prise quand ce test est vérifié; sinon, l'algorithme passe à l'itération suivante, qui commence à partir de l'étape d'évaluation.
\end{itemize}
\subsection{Algorithmes génétiques}
Les algorithmes génétiques (AG) furent initialement développés par J.Holland au début des années 1970 pour l'imitation de certains des processus observés dans l'évolution naturelle. Ce sont des algorithmes d'optimisation reposant sur des techniques qui proviennent de la génétique et de l’évolution naturelle: croisements, mutations, sélection, etc. Leurs bases théoriques furent exposées par Goldberg en 1994 et c'est grâce à son livre que nous devons leur popularisation. Ils essaient de faire une simulation du processus d’évolution des espèces dans leur milieu naturel: soit une transposition artificielle de concepts basiques de la génétique et des lois de survie énoncés par Darwin.\smallskip

Lerman et Ngouenet définissent un algorithme génétique \cite{vallee2004presentation} par:
\begin{itemize}
\item Individu/chromosome/séquence: une solution potentielle du problème.
\item Population: un ensemble de chromosomes ou de points de l’espace de recherche.
\item Environnement: l’espace de recherche.
\item Fonction de fitness: la fonction - positive - que nous cherchons à maximiser.
\end{itemize}
\medskip

Le fonctionnement des AG \cite{guenounou2009methodologie} est extrêmement simple, il se base sur les étapes suivantes:
\begin{enumerate}
\item \textbf{Initialisation:} une population initiale de taille N chromosomes est tirée aléatoirement.
\item \textbf{Évaluation:} chaque chromosome est décodé puis évalué (fitness).
\item \textbf{Reproduction:} création d’une nouvelle population de $N$ chromosomes par l’utilisation d’une méthode de sélection appropriée.
\item \textbf{Opérateurs génétiques:} croisement et mutation de certains chromosomes au sein de la nouvelle population.
\item Retour à la phase 2 tant que la condition d’arrêt du problème n’est pas satisfaite.
\end{enumerate}
\medskip

Leurs champs d’application \cite{vallee2004presentation} sont très larges. Outre l’économie, ils sont employés pour l’optimisation de fonctions (De Jong (1980)), en finance (Pereira (2000)), en théorie du contrôle optimal (Krishnakumar et Goldberg (1992), Michalewicz, Janikow et Krawczyk (1992) et Marco et al. (1996)), ou encore en théorie des jeux répétés (Axelrod (1987)) et différentiels (Özyildirim (1996, 1997) et Özyildirim et Alemdar (1998)). Leur simplicité et efficacité est la raison de ce grand nombre.
\section{Les cartes auto-organisatrices}
Plusieurs études ont été menées sur les modèles de représentation cognitifs appliqués aux systèmes perceptifs tels que les systèmes visuels ou auditifs. Ces études ont permis de mettre en exergue des architectures neuronales dédiées et caractérisées par une auto-organisation naturelle des connexions synaptiques suivant une répartition respectant l'aménagement topologique des régions sensitives du corps humain. Ces études ont stimulé, de cette façon, l'introduction des réseaux de neurones formels, intitulés «cartes auto-organisatrices» \cite{boudjemai2006reconstruction}.\bigskip

Ces cartes auto-organisatrices furent inventées par T.Kohonen en 1982, c'est un type de réseau de neurones artificiels dont l'apprentissage se produit de façon non supervisée. Elles sont influencées par le principe neuronal du cerveau des mammifères et en recourant à certains principes fondamentaux du traitement de l'information à l'intérieur du cerveau, elles permettent de modéliser quelques fonctionnalités essentielles du cerveau. Leur fonction principale est de faire une projection non linéaire des données de haute dimension sur un espace de faible dimension. Les cartes auto-organisatrices sont beaucoup utilisées dans la classification de données \cite{hajjar2014cartes}.\bigskip

Ces réseaux sont constitués d’une grille de neurones (ou nœuds, ou unités) auxquels seront présentés des stimuli. Un stimulus est un vecteur, de dimension $\emph{d}$, décrivant un objet à classer. Ce vecteur est aussi bien apte à être une description des caractéristiques physiques des objets stimuli qui fait référence à des caractéristiques telle que la présence ou l’absence d’un mot-clé dans un document. Toute unité de la grille est reliée au vecteur d’entrée (stimulus) par l’intermédiaire de $\emph{d}$ synapses de poids $\emph{w}$ (Figure 2.4). En réalité, à chaque unité est associé un vecteur de dimension $\emph{d}$ qui contient les poids $\emph{w}$ \cite{rousset1999applications}.\bigskip

\begin{figure}[H]
\centering
\resizebox{12cm}{!}{\includegraphics[]{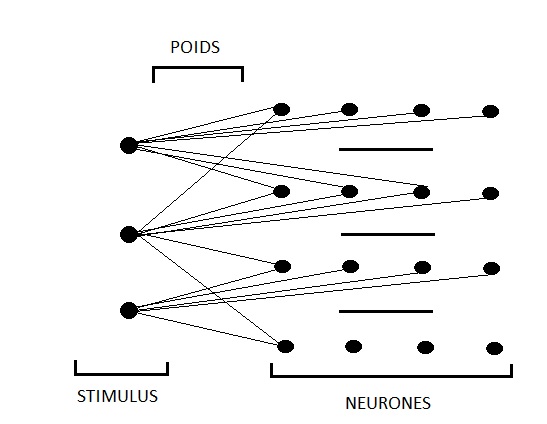}}
\captionof{figure}{Schéma d'une carte de kohonen connectée à un stimulus \cite{rousset1999applications}}
\end{figure}
\section{Clustering}
\subsection{Définition du clustering}
Le clustering, en français regroupement ou partitionnement, est une méthode fondamentale d’analyse de données. C'est une tâche dont l’objectif est de regrouper des objets (documents) similaires. Mais elle diffère de la catégorisation en ce que les documents sont regroupés à la volée au lieu d’utiliser des sujets prédéfinis. Ces objets sont décrits par des caractéristiques, encore appelées attributs, qui décrivent les propriétés des objets. Les groupes recherchés, communément appelés des clusters, forment des ensembles homogènes d’objets du jeu de données partageant des caractéristiques communes. Chaque cluster issu du processus de regroupement doit vérifier les deux propriétés suivantes:
\begin{itemize}
\item La cohésion interne (les objets appartenant à ce cluster soient les plus similaires possibles).
\item L'isolation externe (les objets appartenant aux autres clusters soient les plus distincts possibles).
\end{itemize}

Un des avantages du regroupement est que les documents peuvent émerger dans de multiples sous-thèmes, ce qui garantit qu’un document utile ne sera pas absent des résultats de recherche. Il est largement utilisé pour la reconnaissance de motifs, l’extraction de fonctionnalités, la quantification vectorielle (VQ), la segmentation d’image, l’approximation des fonctions et la fouille de données.\bigskip

En tant que technique de classification non supervisée, le regroupement identifie certaines structures essentielles présentes dans un ensemble d’objets sur la base d’une mesure de similarité. Les méthodes de regroupement peuvent être fondées sur l’identification du modèle statistique \cite{mclachlan1988mixture} ou sur l’apprentissage concurrentiel. Le regroupement est une technique qui n’a pas d’étiquettes de classe prédéfinies, mais en employant des mesures de similarité entre différents objets, elle met les objets les plus similaires dans une classe et les objets dissimilaires dans une autre classe. Le choix de la mesure de similarité permettant de comparer les objets entre eux, va induire la façon de les regrouper \cite{du2010clustering}.\bigskip

De nos jours, le clustering est très utilisé dans plusieurs domaines, comme la vision artificielle, la biologie, l'internet, la reconnaissance des formes, la recherche de documents, la fouille de données, etc. Dans ce contexte, nous allons citer quelques exemples d'applications \cite{eschrich2003fast} du clustering:

\begin{itemize}
\item La réduction de la dimension des bases de données afin de conserver le maximum d'informations utiles dans un espace de dimension inférieure.
\item La prospection de Web (Web mining) et l'analyse des données textuelles (Text Mining) pour la recherche d'information à partir de certains mots clés.
\end{itemize}
\subsection{Processus du clustering}
Étant donné un ensemble d'objets $ X={x_{1},x_{2},...,x_{n}} $  dans l'espace d'attributs  $\Re ^{d}$ avec, $\emph{d}$: dimension de l'espace, $\emph{n}$: le nombre d'objets.\smallskip

$x_{i}=(x_{i1},x_{i2},...,x_{id})$  représente le $i^{éme}$ objet ; et $x_{ij}$  correspond à la valeur du $j^{éme}$  attribut pour le $i^{éme}$  objet. Le but principal du clustering est la recherche des structures similaires dans l'espace d'objets $\Re ^{d}$. Toutes les techniques de clustering suivent le même principe général qui consiste à maximiser la similarité des objets à l'intérieur d'un cluster, et minimiser la similarité des objets entre les clusters, la figure 2.1 est un exemple qui illustre les différentes étapes d'une tâche de clustering \cite{bouguessa2004approche}.

\begin{figure}[H]
\centering 
\resizebox{\linewidth}{!}{\includegraphics[]{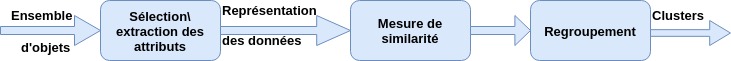}}
\captionof{figure}{Étapes d'une tâche de clustering \cite{bouguessa2004approche}}
\end{figure}

\begin{itemize}
\item La sélection/ extraction des attributs correspond à l'utilisation d'une ou plusieurs transformations des attributs fournis en entrée afin de sélectionner le sous-ensemble le plus efficace à utiliser pour le clustering.
\item La représentation des données se réfère à la spécification du nombre de données, ainsi que la dimension et le type des attributs disponibles pour l'algorithme de clustering.
\item La mesure de similarité consiste à définir une métrique appropriée au domaine des données. Différentes mesures de similarité ont été utilisées dans le clustering. La distance euclidienne est l'une des métriques les plus utilisées.
\item Le regroupement consiste en la construction des groupes similaires, qui représentent le résultat du processus de clustering. Ce résultat peut être dur «hard» ( partition des objets en groupes distincts), ou flou «fuzzy» ( chaque objet a un degré variable d'appartenance à chacun des groupes formés). 
\end{itemize}

\subsection{Clustering flou}
Le regroupement flou généralise les méthodes de classification par partition (telles que k-means et medoid) en permettant à un individu d'être partiellement classé en plus d'un cluster. Dans le clustering classique, chaque individu est membre d'un seul cluster. Supposons que nous avons $K$ clusters et nous définissons un ensemble de variables $m_{i1}$,$m_{i2}$,...,$m_{ik}$ cela représente la probabilité que l'objet $i$ soit classé dans le groupe $k$. Dans les algorithmes de clustering de partition, l'une de ces valeurs sera un et le reste sera zéro. Cela représente le fait que ces algorithmes classent un individu dans un et un seul cluster.\bigskip

Dans le clustering flou, l'appartenance est répartie entre tous les clusters. Le $m_{ik}$ peut maintenant être entre zéro et un, avec la stipulation que la somme de leurs valeurs est un. Nous appelons cela une fuzzification de la configuration du cluster. Il a l'avantage de ne pas forcer chaque objet dans un cluster spécifique. Il a le désavantage qu'il y a beaucoup plus d'informations à interpréter.

\addcontentsline{toc}{section}{Conclusion}
\section*{Conclusion}
Durant ce chapitre nous nous sommes intéressés à la présentation de la fouille de données, la fouille de texte et la recherche d'information. Nous avons également présenté le soft computing et ses composantes. Le prochain chapitre sera dédié à une étude de l’état de l’art.

\chapter{Revue sur la détection de sujets et la théorie de résonance adaptative}
\addcontentsline{toc}{section}{Introduction}
\section*{Introduction}
Au cours des dernières années, de nombreux travaux de recherche sur la détection de sujets ont été menés. Ces approches de détection de sujet peuvent être classées en supervisées et non supervisées. Les approches supervisées nécessitent des experts du domaine pour l'apprentissage des documents de texte sur des sujets conceptuels prédéfinis, et la prédiction sur les étiquettes de sujets peut alors être faite sur des objets de données inconnus. D'un autre côté, les approches non supervisées regroupent des documents texte en différents groupes selon la similarité de leur contenu sans impliquer des experts de domaine dans le but de récupérer des documents texte de sujets identiques ou similaires.\smallskip

Dans ce chapitre, nous allons, dans un premier temps, présenter la détection de sujets. Nous introduirons, dans un deuxième temps, les principales approches de détection de sujets existantes dans la littérature. Nous réservons la dernière partie de ce chapitre à la présentation du réseau ART et des principaux type de ART.

\section{Détection de sujets}

L'indexation de sujet de document, ou, comme on l'appelle parfois, identification de sujet de document, est généralement utilisée pour désigner la tâche de recherche de sujets pertinents pour un ensemble de documents d'entrée. Elle est utilisée dans de nombreuses applications réelles différentes, telles que l'amélioration de la récupération de documents de bibliothèque relatifs à un sujet spécifique. Elle pourrait également être utilisée pour améliorer la pertinence des résultats des moteurs de recherche, en catégorisant les résultats de recherche en fonction de leur sujet général et en donnant aux utilisateurs la possibilité de choisir le domaine qui correspond le mieux à leurs besoins. Il existe différentes tâches dans la fouille de texte qui relèvent de l'indexation des documents, notamment le marquage de documents et l'extraction de phrases clés. Il existe néanmoins plusieurs méthodes d'identification de sujet.

\section{Les approches de l'état de l'art} 

\subsection{A Comparative Study of Topic Identification on Newspaper and E-mail: Bigi 2001}

La quantité de données textuelles disponibles augmentant de manière exponentielle, le traitement automatique de ces données est devenu essentiel. Ce traitement automatique peut être réalisé en utilisant l'identification de sujet (TID).
Cet article \cite{bigi2001comparative} traite du problème d'évaluation des algorithmes TID sur deux types de corpus textuels: les journaux et les e-mails. Une étude comparative de plusieurs méthodes TID est présentée. Cependant, cette étude traite des méthodes classiques de catégorisation des textes. Dans cet article, les méthodes étudiées sont principalement issues de la modélisation statistique des langages dans le domaine de la reconnaissance vocale, donc différentes des méthodes de catégorisation des textes.
\subsubsection*{Démarche et résultats}
Cinq méthodes statistiques impliquées dans les expériences TID ont été utilisées. Les cinq méthodes sont les suivantes:
\begin{itemize}
\item \textbf{Le modèle de langage unigramme:} Le modèle de langage unigramme du sujet est l'un des plus classiques et des plus standards. Il est basé sur un décompte du nombre d'occurrences de chaque mot pour chaque sujet et implique tous les mots de chaque vocabulaire.  Avec ce modèle, les mots non liés à un sujet ont la même importance que ceux qui sont spécifiques à un sujet (mots-clés).
\item \textbf{Le modèle de cache:} Le modèle de cache est basé sur un ensemble de mots-clés sélectionnés automatiquement pour chaque sujet. Ces mots, appelés mots-clés de sujet, ont une distribution statistique obtenue à partir des corpus d'apprentissage.
\item \textbf{Le classifieur TFIDF:} Le classifieur TFIDF représente des sujets en tant que vecteurs. Chacun se caractérise par un ensemble de mots distincts $D_{j}=(w_{j1}, w{j2},…, w_{jn})$ où n est le nombre de mots du sujet j et $w_{jk}$ est le poids du k iéme mot. $w_{jk}$ est défini comme $w_{jk}= nf_{jk}.idf_{k}$, où $nf_{jk}$ est la fréquence du terme, c’est-à-dire le nombre de fois que le mot $w_{k}$ apparaît dans le sujet j. Soit $DF_{k}$ le nombre de documents dans lesquels le mot w apparaît et $|D|$ le nombre total de documents. La similarité entre un sujet j et un document représenté par un vecteur $D_{i}$ est mesurée par le cosinus suivant:\\
\begin{center}
$sim(D_{j}, D_{i})= \frac{\sum_{k=1}^{n}w_{jk}w_{ik}}{\sqrt{\sum_{k=1}^{n}(w_{jk})^{2}}\sum_{k=1}^{n}(w_{ik})^{2}}$
\end{center}
Le sujet sélectionné est celui ayant la plus grande similarité. 
\item \textbf{Modèle basé sur la perplexité:} La perplexité est une mesure issue de la théorie de l'information qui est largement utilisée dans la reconnaissance de la parole, mais rarement dans le TID. La perplexité reflète la capacité d'un modèle de langage à modéliser un texte. Elle est calculée comme la moyenne géométrique inverse de la probabilité d'un texte:\\
\begin{center}
$PP(w_{1}^{N})=\left (P(w_{1})=\prod_{k=2}^{N}P(w_{k}|w_{k-1},...,w_{k-n+1})\right )^{\frac{-1}{N}}$ 
\end{center}
Un bon modèle de langage attribue une faible perplexité aux textes réels. Dans un framework TID, un modèle de langage est créé pour chaque sujet. Ensuite, la perplexité correspondant à chaque sujet est évaluée sur le texte à classer. Le sujet correspondant à la perplexité la plus basse est celui assigné au texte.
\item \textbf{Modèle de poids:} Ce modèle est spécifiquement conçu pour gérer les données bruyantes et éparses. En tant que modèle Cache, il calcule les distances entre le texte et les rubriques. Chaque sujet est représenté par un poids d'unigramme et un mot. Un poids de sujet est attribué à un mot en fonction d'une fonction inversement proportionnelle au nombre de vocabulaires de sujet dans lesquels ce mot est présent.
Dans ce modèle, un score est calculé pour un texte $W_{1}^{N}$ et pour chaque sujet j:\\
$T_{j}(W_{1}^{N})=\beta_{j} \frac{\sum_{k=1}^{N} LP(w_{k}|T_{j}) \eta (w_{k})}{N}$, le sujet résultant est celui correspondant à la plus grande valeur de $T_{j}(W_{1}^{n})$.
\end{itemize}

Deux corpus sont utilisés dans cet article: le premier corpus est constitué de quatre ans (1987-1991) du journal français Le Monde (plus de 80 millions de mots).Le deuxième corpus est constitué de 5 000 e-mails fournis par la start-up française MIC2. 
Chacune des cinq méthodes a d'abord été évaluée sur un journal, puis sur le corpus de courrier électronique. En raison de l'ambiguïté de certains paragraphes, ils ont réalisé deux expériences différentes sur des corpus de journaux. Le premier concerne les paragraphes ayant une seule étiquette (notée {$t_{1}=t_{2}$}) et le second concerne les paragraphes ayant deux étiquettes différentes (notées {$t_{1}, t_{2}$}).
Les résultats sur les paragraphes ayant une seule étiquette montrent que, sur les grands corpus, le modèle de cache fournit les meilleurs résultats (82\% de correcte TID). Néanmoins, les performances des méthodes de base telles que unigramme ou la perplexité (respectivement 79\% et 80\%) sont comparables à celles du modèle de cache.

Dans cet article, cinq méthodes pour le TID ont été présentées. Ces méthodes présentent un avantage tangible lorsqu'elles sont appliquées à deux corpus différents: articles de journaux et courriers électroniques. Il est également important de noter que d'un côté, les articles de journaux décrivent une très grande variété de faits avec un très grand vocabulaire. D'un autre côté, les corpus d'e-mails décrivent une très petite variété de sujets avec un vocabulaire restreint, souvent insuffisants pour apprendre correctement les modèles statistiques.
En général, il y a beaucoup de recherches en cours sur la comparaison des performances de différentes techniques de catégorisation de textes dans des aspects réels en utilisant plusieurs ensembles de données. Les résultats publiés ont montré différentes évaluations des classificateurs en utilisant différentes collections de données. Ces résultats ont indiqué que la performance des techniques évaluées dépend fortement de la collection de données.

\subsection{Topic Detection in Noisy Data Sources: Denecke 2010}
La nécessité de disposer de méthodes permettant l'analyse et l'interprétation automatiques des données Web devient de plus en plus importante. Cela est dû à l’augmentation des données non structurées disponibles. Pour de nombreuses applications qui s'appuient sur des données Web, la détection de sujets de documents à différents niveaux de granularité est nécessaire (niveau document, niveau paragraphe, niveau phrase). Dans cet article \cite{denecke2010topic}, les auteurs ont mis l'accent sur la détection de sujets au niveau de la phrase.
Un algorithme fréquemment cité pour la détection de sujet est la modélisation de sujet par l’allocation latente de Dirichlet (LDA). Jusqu'à présent, il a été principalement appliqué à la détection de sujets au niveau des documents. Le auteurs de cet article l'ont appliqué à la détection des sujets au niveau de la phrase pour les articles de blog et en étudiant sa qualité sur les phrases.
\subsubsection*{Démarche et résultats}
Dans cet article, les auteurs considèrent le thème d’une phrase comme «sujet» qui peut être décrit par un ensemble de mots qui ne doivent pas nécessairement être mentionnés explicitement dans une phrase. Ils déterminent ces sujets en décrivant des termes (appelés «termes de sujet») en utilisant LDA. L'implémentation originale de Blei de LDA est étendue par un algorithme de détection de phrase pour l'appliquer aux phrases. Le processus utilisé pour ce faire est le suivant:
\begin{itemize}
\item \textbf{Répartition des documents:} Premièrement, chaque article est divisé en phrases en utilisant la bibliothèque de partitionnement de phrases fournie par Lingpipe.
\item \textbf{Normalisation des phrases et des mots:} Dans un deuxième temps, les phrases sont normalisées: seuls les noms et les noms propres sont racinisé et conservés pour un traitement ultérieur. Pour détecter des classes de mots et effectuer la racinisation, le Stanford NLP Toolset est utilisé.
\item \textbf{Détection de sujets:} Dans une troisième étape, les sujets ainsi que leurs probabilités sont identifiés pour chaque phrase à l'aide de l'algorithme LDA et sur la base de la représentation vectorielle des phrases (normalisées). Par conséquent, un sujet est décrit par un ensemble de mots dérivés des documents où, à chaque mot, une probabilité est attribuée pour indiquer la pertinence de ce mot pour le sujet. De cette façon, tous les sujets sont décrits par les mêmes mots, mais avec des valeurs de probabilité variables pour chaque mot. La sortie de LDA est finalement la probabilité de chaque mot pour un sujet et la probabilité de chaque sujet pour une phrase.
Pour exécuter LDA, le nombre de clusters à former doit être fixé à l'avance. LDA nécessite également de fixer deux autres paramètres: Le paramètre $\alpha$ qui détermine la dominance d'un sujet dans un document. Et l'hyperparamètre $\beta$ qui peut être interprété comme le nombre d'observations antérieures sur le nombre de fois que les mots sont échantillonnés à partir d'un sujet avant qu'un mot du corpus ne soit observé.
\item \textbf{Sélection de sujets:} Dans une dernière étape, les probabilités déterminées par LDA sont utilisées pour sélectionner le sujet et les termes du sujet pour une phrase. La probabilité par sujet et phrase calculée par LDA indique dans quelle mesure la phrase appartient au sujet.
\item \textbf{Choix des termes de sujets:} LDA fournit également les probabilités de chaque mot pour un sujet. Pour évaluer la qualité de la détection des sujets LDA, les auteurs ont besoin de termes décrivant chaque sujet. Leurs études ont montré que les cinq mots les plus probables convenaient le mieux pour décrire le sujet d'une phrase. Par conséquent, ils considérent les cinq mots avec la plus grande probabilité pour un sujet comme «termes du sujet». Ils sont utilisés pour caractériser un sujet. 
\end{itemize}

LDA modifié est appliqué à trois ensembles de données différents. Le premier ensemble de données (dénommé WebMD) et collecté sur la page Web de WebMD. Le deuxième ensemble de données (appelé Slashdot) comprend des commentaires du site Web Slashdot. Le troisième jeu de données (appelé Reviews) est donné par des avis sur 14 produits collectés sur Amazon. 
Précision de la détection de sujets: En fonction du nombre de sujets et du jeu de données, des valeurs de précision comprises entre 54,2\% et 89,4\% sont atteintes.

Dans cet article, les auteurs ont exploité L'allocation Latente de Dirichlet(LDA) pour déterminer les sujets dans les phrases bruyantes et courtes des blogs. Ils ont effectué une évaluation expérimentale, avec trois ensembles de données du monde réel et ils ont démontré l’applicabilité de la solution proposée qui a donné des résultats très prometteurs. Ces résultats montrent que LDA peut déterminer avec succès des sujets même pour des phrases courtes et bruyantes.
\subsection{A Graph Analytical Approach for Topic Detection: Sayyadi 2013}
Au cours de la dernière décennie, la détection de sujets et la modélisation de sujets ont attiré d'importants travaux de recherche dans plusieurs communautés.
Un événement ou un sujet est défini comme un événement se produisant à un moment et à un endroit précis. Dans cet article, "événements" et "sujets" peuvent être utilisés de manière interchangeable. L'importance croissante des médias sociaux a entraîné un regain d'intérêt pour les méthodes évolutives permettant de traiter de grandes collections bruyantes. L'évolution du sujet apporte une sophistication supplémentaire dans le traitement des défis TDT(topic detection and tracking) sur des flux de données temporelles, en accordant une attention particulière aux thèmes émergents, au changement de sujet, etc.
\subsubsection*{Démarche et résultats}
La modélisation thématique représente les sujets comme un mélange infini sur les distributions de probabilités des ensembles de mots. Le but des auteurs dans cet article est de tenir compte de la cooccurrence des mots dans les documents également.
KeyGraph \cite{sayyadi2013graph} se concentrera sur le défi de maximiser la cooccurrence des mots seulement. Le problème devient alors presque identique à la détection de communauté sur le graphe de cooccurrence des mots-clés, ou KeyGraph. Les auteurs de cet article utilisent l'étiquette KeyGraph pour référencer à la fois le graphe et l'algorithme. KeyGraph comprend les phases suivantes:
\begin{itemize}
\item \textbf{Construction du KeyGraph:} Dans la première phase, ils construisent un graphe de cooccurrence par mot-clé appelé KeyGraph. KeyGraph a un nœud pour chaque mot-clé du corpus.  Les arêtes de KeyGraph représentent la cooccurrence des mots-clés correspondants et sont pondérées par le nombre de cooccurrences.
\item \textbf{L'extraction des caractéristiques du sujet:} La deuxième phase de l'algorithme comprend la détection des communautés et l'extraction des caractéristiques des sujets, c'est-à-dire que les mots clés de chaque communauté seront utilisés pour représenter un sujet.
\item  \textbf{L'attribution de sujets aux documents:} La troisième phase de l'algorithme utilise les caractéristiques des sujets pour assigner des sujets aux documents. Enfin, pour chaque paire de sujets, s'il y a plusieurs documents qui sont assignés aux deux sujets, ils supposent qu'il s'agit de sous-sujets du même sujet parent, et fusionnons ces sujets.
\end{itemize}
KeyGraph a une précision similaire à celle des algorithmes de référence du jeu de données bien connu TDT4, qui est un jeu de données petit et bien annoté. KeyGraph est également comparé à LDA-GS, la modélisation de sujets basée sur LDA en utilisant Gibbs Sampling sur le jeu de données des médias sociaux Spinn3r. KeyGraph a une meilleure précision et un rappel similaire. De plus, la distribution des probabilités du sujet de KeyGraph permet de différencier les documents pertinents et non pertinents et de mieux sélectionner les caractéristiques. Le temps de fonctionnement de KeyGraph est également nettement inférieur à celui de LDA-GS pour les grandes collections.

\subsection{Topic Identification of Arabic Noisy Texts Based on KNN: Abainia 2015}
La diffusion d'informations numériques disponibles sur différents supports numériques conduit à la croissance de données textuelles, ce qui a incité plusieurs chercheurs à contribuer et à étudier dans le domaine de la catégorisation automatique des textes et de l'extraction des connaissances. Dans cette enquête \cite{abainia2015topic}, les auteurs ont abordé le problème de l’identification des textes arabes bruyants.
\subsubsection*{Démarche et résultats}
Deux schémas d'identification de sujets utilisant le classificateur K plus proches voisins(KNN) sont utilisés:
\begin{itemize}
\item \textbf{Basé sur le profil Tf-Idf:} Dans cette approche, pour chaque sujet, ils concatènent N documents d'apprentissage ensemble dans un seul document, pour avoir M documents correspondant aux M sujets. Ensuite, ils calculent les poids Tf-Idf en utilisant les gros documents. Enfin, ils créent N histogrammes correspondant à N documents d'apprentissage de chaque sujet, et ils remplacent les fréquences d'histogramme par des poids Tf-Idf calculés précédemment.
\item  \textbf{Exemple basé sur Tf-Idf:} Dans cette seconde approche, chaque document est pris indépendamment et les poids Tf-Idf sont calculés à l'aide de N * M documents.
\end{itemize}
Le classificateur KNN sert à calculer la distance entre le document d'entrée et N documents d'apprentissage de chaque sujet. Ensuite, ils définissent la valeur de confiance d'avoir le document d appartenant au sujet T.

Ils ont construit un jeu de données en arabe qu'ils ont appelé "ANTSIX" contenant 6 thèmes.
Les distances Cosine et Bray Curtis n'apportent pas de bons résultats avec le profil basé sur Tf-Idf, mais leur précision est fortement accrue avec l'approche basée sur l'exemple de Tf-Idf.
Les mesures de distance proposées FreqSum1 et FreqSum2 fournissent de meilleurs résultats que les autres, à l'exception de la similarité d'intersection d'histogramme. De plus, ils semblent être robustes, simples et peu coûteux en temps de calcul. Enfin, la distance DotProd semble être la meilleure distance utilisée, fournissant la plus grande précision dans les deux approches de KNN (95\% avec une approche basée sur le profil et 95,83\% avec une approche basée sur des exemples).
De plus, nous remarquons que l'approche de l'exemple basé sur Tf-Idf, donne de meilleurs résultats que l'approche du profil basé sur Tf-Idf.
Le classificateur K-plus proches voisins a été utilisé pour traiter le problème d'identification du sujet, dans lequel les auteurs ont proposé deux schémas différents pour former le classifieur (Tf-Idf basé sur le profil et approche basée sur l'exemple Tf-Idf). Ils ont également proposé plusieurs distances associées au KNN.
 
\subsection{Topic Detection and Tracking in News Articles: Patel 2017}
La détection et le suivi des sujets est un sujet difficile dans la recherche d'information qui peut être utilisé dans le processus de fouille de texte.  A partir de données qui ne sont pas structurées dans le fouille de texte, les auteurs de ce papier extraient des informations qui étaient auparavant inconnues. 
L'approche présentée dans ce papier \cite{patel2017topic} combine une variété de techniques d'apprentissage. La détection de sujet est une tâche non supervisée et le suivi de sujet est une tâche supervisée. Ils utilisent le clustering agglomératif pour créer des clusters de thèmes et le classificateur KNN pour le suivi des thèmes. Pour identifier les nouvelles les plus importantes, ils identifient les clusters qui entrent dans la même catégorie.
\subsubsection*{Démarche et résultats}
La figure 2.1 illustre la démarche de l'approche de Patel et al.
\begin{figure}[H]
\centering
\resizebox{\linewidth}{!}{\includegraphics[]{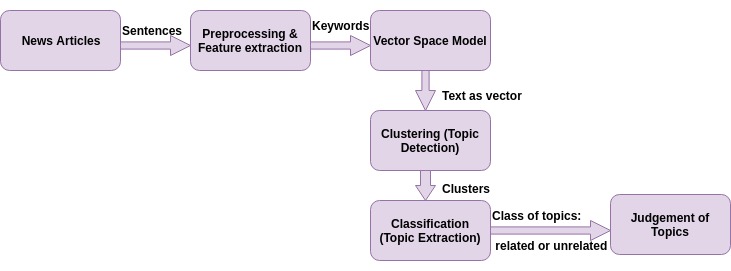}}
\captionof{figure}{Architecture du système proposé par Patel et al. [37]}
\end{figure}
Le prétraitement se déroule en trois étapes. Tout d'abord, la tokenisation s'appliquera aux textes d'articles de presse. Ici, dans la tokenisation, les phrases seront décomposées en mots. Ensuite, de l'ensemble des mots, les mots d'arrêt seront supprimés. Ensuite, la racinisation sera appliquée sur les mots pour obtenir le mot racine.\bigskip

\textbf{Le modèle d'espace vectoriel:} Dans ce travail, ils ont affaire à des textes. Pour trouver des similitudes entre les textes, ils convertissent les textes en vecteurs, pour ce faire ils utilisent le modèle SVM. Ce modèle(SVM) est utilisé comme système de base parce qu'il est facile à mettre en œuvre, robuste et plus compétitif que les autres systèmes élaborés. Pour pondérer le terme, ils utilisent TFIDF et pour la similarité, ils utilisent la mesure de similarité par cosinus.\bigskip

\textbf{Le TFIDF, une approche de pondération des termes:} Les poids des mots trouvés dans les documents sont attribués par le poids du TFIDF. Le mot avec un score tf*idf plus élevé a plus d'importance dans le document. TF représente le nombre de fois que ce terme apparaît dans un article. La IDF représente le nombre de documents contenant ce terme.\bigskip

\textbf{La similarité par cosinus:} Pour mesurer la similarité de deux articles, la similarité par cosinus est utilisée dans le SVM. Il mesure le cosinus de l'angle entre deux vecteurs dans l'espace N-dimensionnel.\bigskip

\textbf{La détection de sujets:} Le clustering agglomératif a été utilisé avec succès pour la détection de sujets. C'est une séquence de partitions imbriquées. Il est défini par un clustering disjoint, qui individualise chacun des N documents au sein d'un cluster. Ce processus est répété et lorsque la séquence progresse en cluster contenant tous les N documents, le nombre de clusters diminue.
Une caractéristique importante de la création de clusters de sujets basés sur des mots-clés est l'occurrence d'un chevauchement de données entre les clusters. Dans cet article les auteurs utilisent le clustering agglomératif avec des mesures de distance moyenne.\bigskip

\textbf{Suivi de sujets:} K plus proches voisins est une méthode de classification basée sur les instances. Le système convertit le document entrant en un vecteur et le compare aux histoires d'apprentissage. Sur la base de la similarité du cosinus entre eux, les auteurs ont choisi les k plus proches voisins. Le score de ce document est calculée en soustrayant les coefficients de similarité des histoires négatives de ceux des histoires positives. Si il est élevé, l'histoire est liée au sujet. Le seuil permettant de déterminer si le score est élevé ou non diffère d'un sujet à l'autre. Il est donc difficile d'appliquer le même seuil à tous les sujets.
\subsection{Topic Identification Method For Textual Document: Jamil 2017}
L'identification d'un sujet est une tâche cruciale pour découvrir la connaissance à partir d'un document textuel. Les méthodes existantes pour l'identification des sujets souffrent d'un problème de comptage de mots car elles dépendent des termes les plus fréquents dans le texte pour produire le mot-clé du sujet. Tous les termes fréquents ne sont pas pertinents. Le présent document \cite{jamiltopic} propose une méthode d'identification des sujets qui filtre les termes importants du texte prétraité et applique un système de pondération des termes pour résoudre le problème des synonymes. 
\subsubsection*{Démarche et résultats}
La méthode d'identification des sujets proposée est illustrée par la figure 2.2.
\begin{figure}[H]
\centering
\resizebox{3cm}{!}{\includegraphics[]{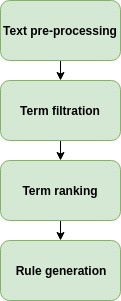}}
\captionof{figure}{Architecture du système proposé par Jamil et al. [32]}
\end{figure}

La méthode se concentre d'abord sur le prétraitement du texte, puis une filtration à terme est effectuée. Un algorithme de filtrage est utilisé pour filtrer les fonctions inutiles du texte et s'assurer que seul le nom est sélectionné. Le nom a été utilisé pour interpréter la structure de la phrase et dépeindre le sujet qui a été discuté par Sagar, Shobha et Kumar (2009). Une fois que les termes ont été filtrés, le classement des termes est effectué lorsque les termes filtrés sont mesurés en examinant le nombre d'occurrences de termes dans le texte.\medskip

Enfin, un algorithme de génération de règles (TopId) est utilisé pour identifier les sujets appropriés pour le texte. L'algorithme détermine quels termes doivent être désignés comme sujet du document texte. L'entrée requise par TopId est le terme pertinent le mieux classé dans le classement des termes.\medskip

Si deux termes ont le même score, TopId prend le premier sujet. En attendant, s'il n'y a pas de correspondance entre les termes et les mots-clés, il est considéré comme hors sujet. Le résultat de TopId sont les règles ainsi que le sujet de chaque couplet. L'état des règles de décision dépend de la disponibilité de certains termes de la base de données de mots-clés. Un classificateur basé sur des règles est utilisé pour déterminer les modèles de termes qui sont liés à différentes classes.\smallskip

La comparaison a été faite en comparant les sujets produits par TopId et Rough Set avec trois experts dans le domaine de recherche du Coran et du Hadith.
Les précisions obtenues par TopId et les trois experts augmentent modérément de 70\% à 78\% et diminuent légèrement à 75\%. Ces résultats ont prouvé que TopId est capable d'identifier des sujets car le nombre total de sujets appariés entre TopId et les trois experts est assez élevé.\smallskip

L'algorithme de génération de règles proposé a permis d'identifier des sujets similaires aux sujets identifiés par les experts. L'algorithme de génération de règles a été conçu pour identifier un sujet basé sur les termes les mieux classés et qui correspond à celui-ci avec les classes thématiques. Les règles générées suggèrent qu'il n'y a qu'un seul terme utilisé pour représenter un sujet pour chaque verset.
\subsection{Bilan et discussion}
Dans cette section, nous allons mener une comparaison théorique entre les approches que nous avons détaillées auparavant. Le tableau 2.1 présente une comparaison théorique entre les approches de détection de sujets. La comparaison est faite suivant les axes suivants:
\begin{itemize}
\item \textbf{Source de données:} cet axe précise les sources de données utilisées dans les systèmes de détection de sujets.
\item \textbf{Techniques utilisées:} cet axe indique les différents techniques utilisées dans les systèmes de détection de sujets.
\end{itemize}
\begin{table}[H]
\centering
\caption{Récapitulatif des travaux de l’état de l’art}
\resizebox{\textwidth}{!}{
\begin{tabular}{|l|l|l|}
\hline
\large \textbf{Approches}                                               & \large \textbf{Sources de données                                                                                            } & \large \textbf{Techniques utilisées                                                                                                                                             } \\ \hline 
Bigi et al. \cite{bigi2001comparative} & \begin{tabular}[c]{@{}l@{}}Le journal français Le Monde\\ E-mails\end{tabular}                                 & \begin{tabular}[c]{@{}l@{}}Modèle de langage unigramme\\ Modèle de cache\\ Classifieur Tf-IDF\\ Le modèle basé sur la perplexité\\ Le modèle pondéré\end{tabular} \\ \hline
Denecke et al. \cite{denecke2010topic} & \begin{tabular}[c]{@{}l@{}}WebMD: blogs de santé\\ Slashdot: commentaires du site web \\Slashdot\end{tabular}    & LDA                                                                                                                                                               \\ \hline
Sayyadi et al. \cite{sayyadi2013graph} & \begin{tabular}[c]{@{}l@{}}TDT4\\ Spinn3r: articles de blog collectés\end{tabular}                             & KeyGraph                                                                                                                                                          \\ \hline
Abainia et al. \cite{abainia2015topic} & \begin{tabular}[c]{@{}l@{}}ANTSIX: collection de textes \\ correspondant aux discussions de forum\end{tabular} & \begin{tabular}[c]{@{}l@{}}Classifieur k-plus proches voisins \\basé sur Tf-Idf\end{tabular}                                                                     \\ \hline
Patel et al. \cite{patel2017topic}     & \begin{tabular}[c]{@{}l@{}}Articles de nouvelles : texte recueilli\\  de diverses sources\end{tabular}         & \begin{tabular}[c]{@{}l@{}}Regroupement agglomératif basé sur \\ la liaison moyenne\\ Similarité par cosinus et K-plus \\ proches voisins\end{tabular}               \\ \hline
Jamil et al. \cite{jamiltopic}         & \begin{tabular}[c]{@{}l@{}}Coran traduit en anglais extrait\\  du site web Surah.my\end{tabular}               & Algorithme de génération des régles                                                                                                                               \\ \hline
\end{tabular}}
\end{table}

Les systèmes de détection de sujets visent à rechercher le(s) sujet(s) traité(s) dans un document. Leur objectif est de regrouper un ensemble de documents suivant un sujet précis. À la lumière de ce qui a été présenté dans ce chapitre et à partir du tableau 2.1, nous pouvons conclure que toutes les approches citées dans la littérature sont différentes aux niveau des techniques utilisées pour effectuer une détection de sujets. Parmi les limites des systèmes de détection de sujets, nous avons constaté qu'ils ne prennent pas en considération les synonymes dans le choix du vocabulaire.
\section{Théorie de la résonance adaptative (ART)}
\subsection{ART}
Les principes tirés d’une analyse des littératures expérimentales dans la vision, la parole, le développement cortical et l’apprentissage par renforcement, y compris le blocage de l’attention et les interactions cognitivo-émotionnelles, ont conduit à l’introduction de la résonance adaptative en tant que théorie du traitement de l’information cognitive humaine \cite{grossberg1976adaptive}. La formulation de cette théorie est basée sur de nombreuses notions issues de la biologie, de la psychologie et des mathématiques.\smallskip

La théorie de la résonance adaptative, ou ART, a évolué comme une série de modèles de réseaux neuronaux en temps réel qui effectuent l’apprentissage, la reconnaissance de motifs et la prédiction non supervisée et supervisée (\cite{duda2001pattern}; \cite{levine2000introduction}). Les modèles d’apprentissage non supervisé comprennent ART 1 \cite{carpenter1987massively} pour les schémas d’entrée binaires et le fuzzy ART \cite{carpenter1991fuzzy} pour les modèles d’entrée analogiques. Les modèles ARTMAP \cite{carpenter1992fuzzy} combinent deux modules non supervisés pour mener à bien l’apprentissage supervisé. De nombreuses variantes des réseaux de base supervisés et non supervisés ont depuis été adaptées aux applications technologiques et aux analyses biologiques \cite{carpenter2011adaptive}.\smallskip

ART est actuellement la théorie cognitive et neuronale la plus développée disponible. Le pouvoir prédictif d’ART est sa capacité à mener à bien un apprentissage rapide, progressif et stable non supervisé et supervisé en réponse à un monde en mutation. Les réseaux ART apprennent les modèles d’entrée en les classant de manière non supervisée: il n’y a pas de professeur externe qui indique au réseau sous quelle catégorie un modèle d’entrée doit être stocké. Le fait que les progrès de l’apprentissage ne soient pas supervisés impose des restrictions sur la façon dont les processus d’entrée sont traités dans les réseaux d’ART. Ces restrictions découlent de la prise en compte du dilemme stabilité-plasticité. Selon ce dilemme, un réseau neuronal doit être plastique, afin de stocker de nouveaux modèles d’entrée; cependant, il devrait également être stable, afin de protéger les motifs stockés d’être effacés (voir Grossberg, 1987b). Les réseaux d’ART font face au dilemme stabilité-plasticité en traitant les nouveaux modèles d’entrée différemment des anciens modèles de saisie.\smallskip

Ci-dessous, nous décrivons brièvement trois notions centrales: (a) la notion de deux étapes, (b) la notion d'entrée à deux composantes, et (c) la notion d'intégration des processus ascendants et descendants.\medskip

\begin{enumerate}[label=(\alph*)]
\item Notion de deux étapes : La théorie se rapporte aux réseaux de deux couches interconnectées, désignées F1 et F2 dans la figure 2.1(a). F1 est une couche d’entrée et F2 est une couche de sortie. Un modèle dans F2 représente une catégorisation d’un motif en F1. Les connexions entre F1 à F2 (représentées par les flèches de la figure 2.3 (a)) permettent aux deux couches de communiquer.

\item La notion d'entrée à deux composantes : Les signaux d’entrée sont composés de deux composants: un composant d’excitation spécifique et un non spécifique (voir la figure 2.3(b)). Le composant d’information contient l’information ou le contenu structuré dans l’entrée ("Qu’est-ce que c’est?"). La composante d’éveil représente l’activation générale qui est générée par la présence de l’entrée seule, quel que soit son contenu ("il y a quelque chose!").

\item La notion d'intégration des processus ascendants et descendants : L’intégration de l’apport de l’environnement et des attentes générées en interne sur la base de la connaissance de l’environnement est une notion importante dans l’ART (voir la figure 2.3 (c)). L’intégration du traitement des signaux d’entrée de bas en haut avec les signaux basés sur le savoir (haut-bas) fournit une clé du dilemme stabilité-plasticité. Il permet aux réseaux d’ART de différencier les modèles récents et anciens: un nouveau modèle ne correspond pas aux attentes tandis qu’un ancien le fait \cite{postma1995adaptive}.
\end{enumerate}
\smallskip

\begin{figure}[H]

  \begin{minipage}[]{0.3\linewidth}
   \centering
   \includegraphics[scale=0.5]{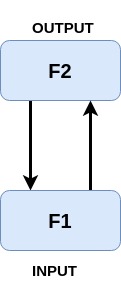} 
   \subcaption{ }     
  \end{minipage} \hfill
  \begin{minipage}[]{0.3\linewidth}
   \centering
   \includegraphics[scale=0.7]{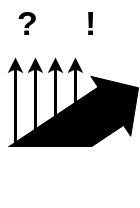} 
   \subcaption{ }     
  \end{minipage}
  \begin{minipage}[]{0.3\linewidth}
   \centering
   \includegraphics[scale=0.5]{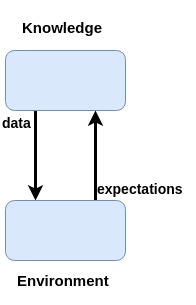} 
   \subcaption{ }     
  \end{minipage}
  \label{fig:ma_fig}
\caption{Les notions basiques de la théorie de résonance adaptative \cite{postma1995adaptive}}
\end{figure}

\subsection{ARTMAP}
Carpenter introduit en 1991 predictive ART ou ARTMAP qui est une nouvelle architecture de réseau neuronal, qui de manière autonome apprend à classer de façon arbitraire un nombre considérable de vecteurs arbitrairement ordonnés en catégories de reconnaissance fondées sur le succès prédictif. Ce système d’apprentissage supervisé est une extension de la théorie de la résonance adaptative (ART). \smallskip

Il est formé à partir d’une paire de modules de théorie de la résonance adaptative ($ART_{a}$ et $ART_{b}$) aptes de se structurer de manière automatique dans des catégories de reconnaissance stable pour répondre à des séquences arbitraires de modèles d’entrée.\smallskip
 
\begin{center}
\begin{figure}[H]
\centering
\resizebox{\linewidth}{10cm}{\includegraphics[]{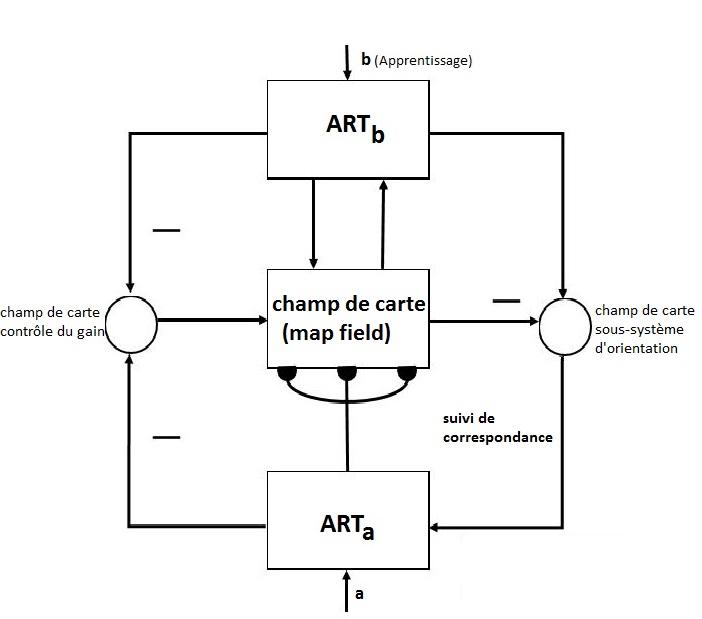}} 
\captionof{figure}{Architecture de ARTMAP \cite{carpenter1991artmap}}
\end{figure}
\end{center}

Les éléments essentiels d’un système ARTMAP sont élucidés par la figure 2.4. Deux modules, $ART_{a}$ et $ART_{b}$ qui permettent de lire les entrées vectorielles \textit{a} et \textit{b}. Si $ART_{a}$ et $ART_{b}$ ont été déconnectés, chaque module auto-organise des groupements de catégories pour les ensembles d’entrée séparés. Codant des vecteurs d’entrée binaires, $ART_{a}$ et $ART_{b}$ sont des modules ART1 à apprentissage rapide.\smallskip

$ART_{a}$ et $ART_{b}$ sont ici connectés par un module inter-ART qui, de plusieurs manières, ressemble à ART1. Ce module inter-ART contient un champ de carte (map field) qui contrôle l’apprentissage d’une carte associative des catégories de reconnaissance $ART_{a}$ aux catégories de reconnaissance $ART_{b}$. Cette carte n’associe pas directement les exemples \textit{a} et \textit{b}, mais associe les représentations compressées et symboliques des familles des exemples \textit{a} et \textit{b}. Le Champ de carte permet également de contrôler le suivi de correspondance du paramètre de vigilance de $ART_{a}$. Une incompatibilité avec le champ de carte entre la catégorie $ART_{a}$ activée par une entrée \textit{a} et la catégorie $ART_{b}$ activée par l’entrée \textit{b} fait augmenter la vigilance de $ART_{a}$ par la somme minimale requise pour que le système recherche et si nécessaire, puisse apprendre une nouvelle catégorie $ART_{a}$ dont la prédiction correspond à la catégorie $ART_{b}$.\smallskip

Ce signal de réinitialisation de la vigilance inter-ART est une forme de «rétro-propagation du gradient» de l’information, mais qui est différente de la rétro-propagation du gradient qui se produit dans le réseau de rétro-propagation du gradient. Par exemple, la recherche lancée par la réinitialisation inter-ART peut attirer l’attention sur un nouveau groupe de fonctions visuelles qui pourraient être intégrées à travers l’apprentissage dans une nouvelle catégorie de reconnaissance $ART_{a}$. Ce processus est comparable à l’apprentissage d’une catégorie de «bananes vertes» en fonction du retour «goût». Cependant, ces événements ne reproduisent pas les caractéristiques de goût dans la représentation visuelle des bananes, comme cela peut se produire en utilisant le réseau de rétro-propagation du gradient. Au contraire, le suivi des correspondances réorganise la façon dont les fonctions visuelles sont regroupées, fréquentées, apprises et reconnues pour prédire un goût attendu \cite{carpenter1991artmap}.
\subsection{Fuzzy ARTMAP}

Le fuzzy ARTMAP est un réseau de neurones artificiels qui appartient à la famille ARTMAP qui permet l’apprentissage supervisé et non supervisé des observations dont les caractéristiques sont définies dans l’espace des nombres réels. La fonction d’appartenance dont la valeur se situe entre 0 et 1 est intégrée par le modèle. Carpenter, Grossberg, Markuzon, Reynolds et Rosen ont introduit le principe de base des réseaux de neurones de type FAM en 1992. Ce type de réseau utilise le réseau fuzzy ART \cite{Carpenter:1991:FAF:125301.125311} qui emploie un algorithme non supervisé pour la génération des groupes (clustering). Il utilise la stratégie de vote qui le mène à faire un apprentissage rapide sur les poids et la reconnaissance des catégories pour un ensemble de données d’apprentissage.\smallskip

Il se distingue également des autres réseaux de neurones par le fait qu’il change ses poids synaptiques après chaque observation, plutôt que de réaliser un apprentissage après avoir inspecté l’ensemble des observations disponibles. Ceci lui donne un certain avantage pour le traitement des applications d’apprentissage dites en-ligne. En réalité, ce type de réseau est obtenu en remplaçant les modules ART1 [\cite{carpenter1987massively},\cite{carpenter1991pattern}] du système ARTMAP par un module fuzzy ART [\cite{carpenter1991fuzzy},\cite{Carpenter:1991:FAF:125301.125311}]. En remplaçant l’opérateur d’intersection de type booléen du module ART1 par un opérateur ET de type flou on obtient le module fuzzy ART.\smallskip

Le FAM est simple à utiliser car un petit nombre de paramètres est indispensable pour sa mise en œuvre. En effet, son comportement est fondé principalement sur le paramètre de choix ($\alpha$), le paramètre de vigilance de base ($\bar{\rho }$), de MatchTracking ($\varepsilon$ ) et le paramètre d’apprentissage ($\beta$). Sa complexité algorithmique est faible et peut être mise en œuvre très efficacement en électronique numérique. Il existe d’autres algorithmes qui sont inspirés de la logique floue, mais le FAM se différencie de ceux-ci par le fait qu’il modifie ses poids synaptiques après chaque observation (apprentissage séquentiel) au lieu de réaliser un apprentissage "batch" après avoir examiné l’ensemble des observations disponibles. Ceci est un avantage remarquable pour les applications d’apprentissage dites en-ligne. La figure 2.5 représente l’architecture du modèle FAM. \smallskip

\begin{center}
\begin{figure}[H]
\centering
\resizebox{\linewidth}{8cm}{\includegraphics[]{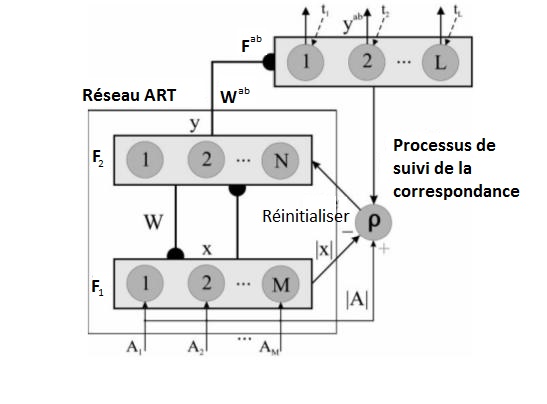}}
\captionof{figure}{Architecture de Fuzzy ARTMAP \cite{Granger2001A}}
\end{figure}
\end{center}

Le réseau fuzzy ARTMAP comprend trois couches distinctes, soit la couche $F_{1}$,$F_{2}$ et $F^{ab}$ (le champ MAP). La valeur des caractéristiques du patron A est propagée à la couche $F_{1}$. La couche $F_{1}$ est reliée à la couche $F_{2}$ par les poids synaptiques W et elle produit le vecteur d’activation $x=A\wedge W$ . La couche $F_{2}$ représente les catégories formées dans le réseau par lequel le patron est classifié. Elle est reliée au champ MAP par les poids $W^{ab}$ et elle produit le vecteur d’activation y. Le vecteur d’activation $y^{ab}$ est produit par le champ MAP. En fait, selon la catégorie de la couche $F_{2}$ sélectionnée, il active une classe K. Au cours de la phase d'apprentissage, le champ MAP active, si besoin, le processus du suivi de la correspondance (MatchTracking), qui emploie les valeurs du patron A et de la couche $F_{1}$(x) dans le but de modifier la recherche de la catégorie adaptée pour le patron présenté \cite{Henniges2006pso}.\smallskip

Les simulations ont élucidé la performance de Fuzzy ARTMAP relativement aux systèmes de propagation de dosages et d'algorithme génétique de référence. Dans tous les cas, les simulations de Fuzzy ARTMAP mènent à des niveaux favorables de précision prédictive, de vitesse et de compression de code dans les paramètres en ligne et hors ligne. Fuzzy ARTMAP est également facile à utiliser . Il a un petit nombre de paramètres, n’exige pas l'élaboration de système spécifique au problème ou de choix de valeurs de poids initiales et ne s'accroche pas au minimums locaux \cite{carpenter1992fuzzy}.
\subsection{Fuzzy ART}
\subsubsection{Définition de Fuzzy ART}

Le système Fuzzy ART introduit ici les calculs incorporés de la théorie des ensembles flous en ART1. C'est une structure de cluster d'apprentissage non supervisé pour les modèles d'entrée à valeur analogique qui a été introduite par Carpenter en 1991. Par exemple, l'opérateur d'intersection ($\cap $) employé dans l'apprentissage ART1 est remplacé par l'opérateur MIN ($\wedge$) de la théorie des ensembles flous. Fuzzy ART se réduit à ART1 en réponse à des vecteurs d'entrée binaires, mais peut également apprendre des catégories stables en réponse à des vecteurs d'entrée analogiques.\smallskip

En particulier, l'opérateur MIN se réduit à l'opérateur d'intersection dans le cas binaire. L'apprentissage est stable car tous les poids adaptatifs ne peuvent que baisser dans le temps. Une étape de prétraitement, nommée codage du complément, utilise des réponses cellulaires et hors cellules pour prévenir la prolifération des catégories. Le codage complémentaire normalise les vecteurs d'entrée tout en gardant les amplitudes des activations des fonctionnalités individuelles \cite{carpenter1991fuzzy}. En général Fuzzy-ART est une architecture de réseau neuronal de clustering qui auto-organise des codes de  reconnaissance pour répondre à des séquences de motifs d'entrée analogiques ou binaires.

\subsubsection{Architecture de Fuzzy ART}
Fuzzy ART a la même structure que le système ART1. Il est composé de deux couches de cellules de calcul ou  de neurones F1 et F2 et d'un sous-système de vigilance contrôlé par un paramètre de vigilance  réglable $\rho \in   [0,1]$.\smallskip

La couche F1 est la couche d'entrée qui se compose de N cellules d'entrée. Chaque cellule d'entrée reçoit  une composante $I_{i} \in [0,1]$ du vecteur d'entrée analogique $I=(I_{1},…,i_{N})$. La couche F2 est la couche de catégorie.\smallskip
 
Elle se compose de M cellules, chacune de ces cellules représente une catégorie possible. Chaque cellule de catégorie reçoit une entrée $T_{j}$. Chaque neurone i appartenant à la couche F1 est relié à chaque neurone j de la couche F2 par une liaison synaptique de poids $z_{ij}^{bu}$. Chaque neurone j appartenant à la couche F2 est relié à chaque neurone i de la couche F1 par une liaison synaptique de force $z_{ji}^{td}$. Dans fuzzy ART $z_{ij}^{bu} =z_{ji}^{td}$. En conséquence, nous ferons référence aux poids comme $z_{ij}= z_{ij}^{bu}=z_{ji}^{td}$. 
Les différences majeurs \cite{serrano2012adaptive} entre les architectures ART1 et Fuzzy ART sont les suivantes :\smallskip

\begin{itemize}
\item Les vecteurs d'entrée sont analogiques. C'est $I=(I_{1},…,I_{N})$ est un vecteur N-dimensionnel avec toute composante $I_{i} \in [0,1]$.
\item Il n'existe qu'un jeu de vecteurs de poids de valeur analogique $z_{j}=(z_{1j},…,z_{Nj}) $;   $j= 1,..,M$.
\item Dans le calcul des fonctions de choix $T_{j}$, de la règle d'apprentissage et du critère de vigilance: l'opération d'intersection $\cap$  (ET binaire) est substituée par l'opérateur flou MIN $\wedge$  (ET analogique).
\end{itemize}

\begin{center}
\begin{figure}[H]
\centering
\includegraphics[width=8.83cm, height=6.23cm]{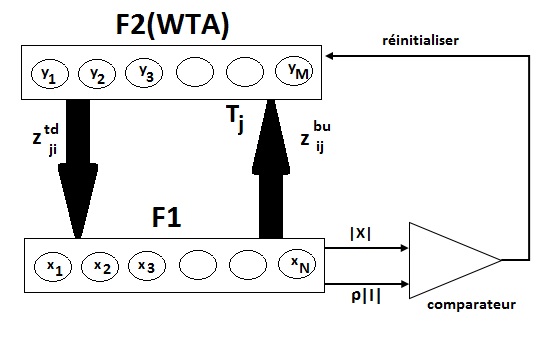}
\captionof{figure}{Structure topologique de l'architecture de Fuzzy ART \cite{serrano2012adaptive}}
\end{figure}
\end{center}
\subsubsection{Résumé de l'algorithme de Fuzzy ART}
Vecteur d'entrée: chaque entrée I est un vecteur dimensionnel $M(I_{1},...,I_{M})$, où chaque composante $I_{i}$ est dans l'intervalle [0,1].\smallskip

Le vecteur de poids: chaque catégorie (j) est définie par un motif qui  est un vecteur de poids $w_{j}\equiv (w_{j1}, ..., w_{jM})$. Le nombre (N) des catégories codées peut être arbitrairement grand.\smallskip

Initialement
\begin{eqnarray}
w_{j1} = . . . = w_ {jM} = 1         
\end{eqnarray}
et chaque catégorie est proclamée non engagée. Un nœud s’engage une fois qu’il a été sélectionné pour le codage. Comme indiqué ci-dessous, chaque composant de poids $w_{ji}$ est monotone non-croissant dans le temps et converge en conséquence à une limite. Le vecteur de poids de Fuzzy ART $w_{j}$ abrite à la fois les vecteurs de poids de bas en haut et de haut en bas de ART 1.\smallskip

Paramètres: Un paramètre de choix $\alpha > 0 $, un paramètre d'apprentissage $ \beta \in \left. [ 0,\right. 1 ] $; et un paramètre de vigilance $\rho  \in \left. [ 0,\right. 1]$ ; déterminent la dynamique de Fuzzy ART.\smallskip

Choix de catégorie: pour chaque catégorie j, la fonction de choix $T_{j}$ est définie par:
\begin{equation}
T_{j}= \frac{ \left | I\wedge W_{j} \right | }{\alpha +\left | w_{j} \right | }      
\end{equation}

Où l'opérateur d'intersection floue (Zadeh, 1965) $\wedge$ est défini par :
\begin{equation}
(x\wedge y)_{i}= min(x_{i}, y_{i})                
\end{equation}

et où la norme $\left  |	~   \right |$ est définie par :
\begin{equation}
\left | x \right |=\sum_{i=1}^{M}x_{i}           
\end{equation}

Le choix de catégorie est indexé par J, où
\begin{equation}
T_{J}=max{~T_{j}:j=1...N}.                                  
\end{equation}

Si plus d'un index j fournit un $T_{j}$ maximum, le nœud avec le plus petit index est sélectionné. Ainsi, les nœuds deviennent engagés dans l'ordre j = 1,2,3 ,. . . .

Résonance ou réinitialisation: si la fonction de correspondance du nœud choisi répond au critère de vigilance alors la résonance se produit. C'est,
\begin{equation}
\frac{\left | I\wedge w_{J} \right |}{\left | I \right |}\geq \rho                     
\end{equation}

L'apprentissage s'ensuit, tel que défini ci-dessous. La réinitialisation de l'incompatibilité se produit si
\begin{equation}
\frac{\left | I\wedge w_{J} \right |}{\left | I \right |}< \rho      
\end{equation}

Ensuite, la valeur de la fonction de choix $T_{J}$ est réinitialisée à -1 pendant la durée de la présentation d'entrée. Un nouvel indice J est choisi, par (2.5). Le processus de recherche continu jusqu'à ce que J satisfasse (2.6).\smallskip

Apprentissage: Le vecteur de poids $W_{J}$ est mis à jour selon l'équation
\begin{equation}
w_{J}^{(new)}=\beta (I\wedge w_{J}^{(old)})+(1-\beta )w_{J}^{(old)}    
\end{equation}

L'apprentissage rapide correspond au réglage $\beta = 1$. La loi d'apprentissage (2.8) est la même que celle utilisée par Moore (1989).\smallskip

Option d'engagement rapide: Il est utile de définir $\beta = 1$ lorsque J est un nœud non engagé, puis de prendre, $\beta <1$ après la validation de la catégorie, pour un codage efficace des ensembles d'entrée bruyants. Alors $w_{J}^{(new)} = I$ la première catégorie J devient active. Cette option d'engagement rapide et de recodage lent correspond à l'apprentissage ART à des taux intermédiaires (Carpenter, Grossberg et Rosen, 1991; Moore, 1989).\smallskip

Option de normalisation des entrées: Moore (1989) décrit un problème de prolifération de catégorie qui peut survenir dans certains systèmes ART analogiques lorsqu'un grand nombre d'entrées érodent la norme des vecteurs de poids. Si les entrées sont normalisées, la prolifération des catégories est évitée dans le Fuzzy ART; c'est-à-dire si
\begin{equation}
\left | I \right |\equiv constant                      
\end{equation}

pour toutes les entrées \emph{I}. En pré-traitant chaque vecteur entrant \emph{a}, la normalisation peut être obtenue, par exemple le réglage
\begin{equation}
I=\frac{a}{\left | a \right |}                                  
\end{equation}

Un autre périphérique, appelé codage du complément, permet une normalisation tout en préservant les informations d'amplitude. La réponse et l'absence de réponse \emph{a} sont représentées par le codage complémentaire. Pour définir un tel code dans sa forme la plus simple, laissez lui-même représenter la réponse. Le complément de \emph{a}, désigné par $a^{c}$, représente la réponse négative, où
\begin{equation}
a_{i}^{c}\equiv 1-a_{i}               
\end{equation}

L'entrée I du système de reconnaissance est le vecteur bidimensionnel
\begin{equation}
I\equiv (a,a^{c})\equiv (a_{1},...,a_{M},a_{1}^{c},...,a_{M}^{c})            
\end{equation}

Notez que
\begin{equation}
\left | I \right |=\left | (a,a^{c}) \right |= M                       
\end{equation}

de sorte que les entrées pré-traitées en forme de codage complémentaire sont automatiquement normalisées. Initialement, $w_{j1}=...= w_{j,2M} = 1$. \cite{carpenter1991fuzzy}

\addcontentsline{toc}{section}{Conclusion}
\section*{Conclusion}
Durant ce chapitre nous nous sommes intéressés à la présentation de la détection de sujets. Nous avons aussi présenté les différentes approches de détection de sujets proposées dans la littérature. Nous avons enfin présenté le réseau ART ainsi que ses principaux types. En se basant sur les notions qui ont été énoncées durant les deux premiers chapitres de ce mémoire, nous pourrons ainsi présenter la nouvelle approche que nous proposons pour le clustering flou basé sur les réseaux de neurones adaptatifs qui fera l'objet du prochain chapitre.

\chapter{ClusART: Une approche de détection de sujets basée sur les réseaux de neurones adaptatifs (ART)}
\addcontentsline{toc}{section}{Introduction}
\section*{Introduction}
Dans ce chapitre, nous présentons une approche de détection de sujets basée sur les réseaux de neurones adaptatifs ART : appelée ClusART. Après avoir étudié l’état de l’art des approches de détection de sujets, nous commençons dans une première partie par la présentation du problème. Par la suite, nous allons présenter ClusART et son architecture. Ensuite, nous allons détailler les phases de ClusART. Enfin, nous allons finir par une conclusion.
\section{Présentation du problème}
L'analyse de cluster ou le clustering consiste à grouper un ensemble d'objets de sorte que les objets d'un même groupe (appelé cluster) se ressemblent d'avantage (dans un sens ou dans un autre) que dans les autres groupes (clusters). C'est une tâche principale de la fouille de données exploratoire, et une technique courante pour l'analyse de données statistiques. Elle est utilisée dans de nombreux domaines, notamment l'apprentissage automatique, la reconnaissance de formes, l'analyse d'images, la recherche d'information, la bio-informatique et l'infographie. Pour effectuer le clustering on peut également utiliser les réseaux de neurones.\bigskip

Les réseaux neuronaux artificiels (RNA) ou les systèmes connexionnistes sont des systèmes informatiques inspirés des réseaux neuronaux biologiques qui constituent le cerveau animal. De tels systèmes apprennent (améliorent progressivement l'exécution) des tâches en considérant des exemples, généralement sans programmation spécifique à une tâche. Par exemple, dans la reconnaissance d'images, ils pourraient apprendre à identifier les images contenant des lapins en analysant des exemples d'images qui ont été étiquetés manuellement «lapin» ou «non lapin» et en utilisant les résultats pour identifier les lapins dans d'autres images. Ils le font sans aucune connaissance a priori sur les lapins, par exemple, qu'ils ont de la fourrure, des queues, des moustaches et des visages félins. Au lieu de cela, ils développent leur propre ensemble de caractéristiques pertinentes à partir du matériel d'apprentissage qu'ils traitent.\medskip

Dans notre cas on s'intéressera à un modèle de clustering flou basé sur les réseau de neurones à architecture évolutive qui est la théorie de résonance adaptative communément appelée réseau ART.\medskip

Les modèles proposés pour le clustering flou des documents basés sur les réseaux de neurones adaptatifs sont encore à un stade précoce, ils présentent ainsi certaines limites. Premièrement, ils nécessitent une quantité très importante de données d’apprentissage afin de détecter une bonne représentation qui établit des alignements appropriés entre le document original et le thème correspondant. 
\section{Présentation de ClusART}

Dans ce travail nous proposons une nouvelle approche de clustering flou basée sur les réseaux de neurones adaptatifs ART (Adaptive Resonance Theory network). Les réseaux ART sont des réseaux de neurones compétitifs, constitués d’un ensemble de neurones et qui sont à la base d’un modèle d’apprentissage non supervisé.\bigskip

Ils sont capables, à partir d’une structure autonome, de développer des clusters stables à partir des entrées arbitraires. En effet ART n’apprend que dans son état de résonance dans lequel un vecteur prototype correspond d'assez près au vecteur d’entrée courant est considéré. Par ailleurs, ART dispose d’une structure d’autocontrôle permettant un apprentissage autonome.
\section{Architecture de ClusART}
L’approche, ClusART, que nous proposons dans ce chapitre et que nous illustrons par la figure 3.1 se déroule en trois phases, à savoir :
\begin{itemize}
\item \textbf{Phase de prétraitement:} Lors de cette phase de prétraitement des données, on procède au nettoyage de ces dernières pour ne plus avoir à faire face aux informations non pertinentes et redondantes présentes ou aux données bruitées et non fiables. Cette phase comprend la tokenisation, la normalisation et la racinisation.
\item \textbf{Phase de création des vecteurs:} Cette phase permet de former les vecteurs correspondants à chaque document des documents pré-traités précédemment.
\item \textbf{Phase de Clustering:} Cette phase prend en entrée les vecteurs résultants de la phase précédente afin de procéder au clustering. Lors de cette phase on a utilisé l'algorithme de FuzzyART qui est un réseau de neurone adaptatif qui implémente la logique floue dans la reconnaissance des formes de ART, améliorant ainsi la généralisabilité. Une caractéristique facultative (et très utile) de FuzzyART est le codage du complément, un moyen d'incorporer l'absence de caractéristiques dans les classifications de modèles, ce qui contribue grandement à prévenir la prolifération inefficace et inutile des catégories. Les mesures de similarité appliquées sont basées sur la norme L1. FuzzyART est connu pour être très sensible au bruit. Lors de cette phase FuzzyART est utilisé pour l'apprentissage. Dans un deuxième temps on a utilisé un classifieur paragraph Vector pour le test.
\end{itemize}

\begin{figure}[H]
\centering 
\resizebox{\linewidth}{!}{\includegraphics[]{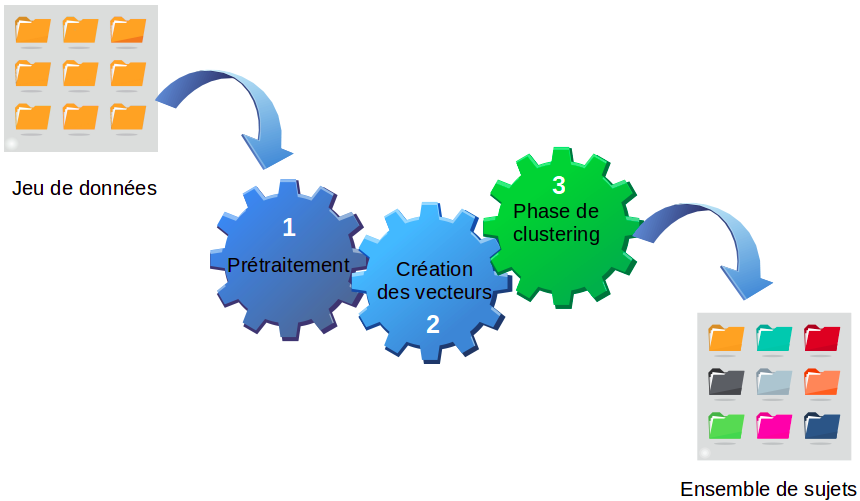}}
\captionof{figure}{Architecture de notre approche ClusART}
\end{figure}

\section{Phase de prétraitement}
Notre collection de données représente des données textuelles qui sont en langage naturel qui est une forme semi-structurée. Il est vraiment difficile d'en extraire des informations utiles et ainsi, ces données ne peuvent pas être utilisées pour la prédiction ou toute autre fonction utile. Ces données doivent être raffinées pour pouvoir être utilisées ultérieurement. Lors de cette phase et après l'extraction de texte, nous passons à la partie la plus cruciale, le nettoyage du texte extrait. Nous avons utilisé plusieurs techniques de prétraitement. Ces techniques éliminent le bruit des données textuelles, identifient plus tard la racine de chaque mot et réduit la taille des données textuelles pour assurer un traitement plus efficace lors des phases suivantes.\bigskip

La figure 3.2 suivante illustre les étapes de prétraitement qu'on va détailler par la suite.\bigskip

\begin{figure}[H]
\centering 
\resizebox{13cm}{!}{\includegraphics[]{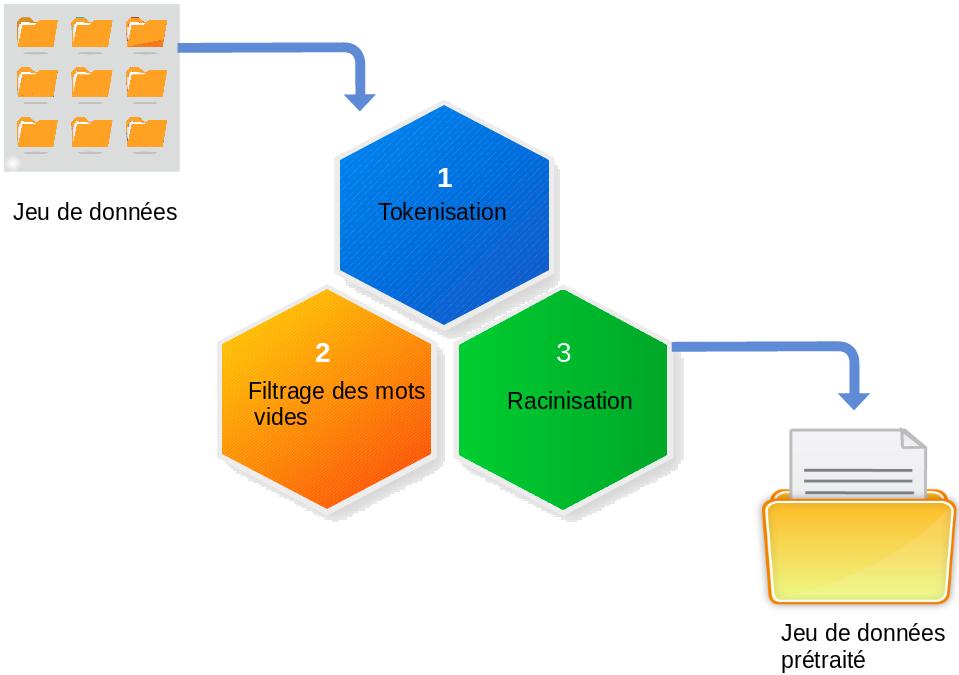}}
\captionof{figure}{Étapes de prétraitement}
\end{figure}

\subsection{Tokenisation}
La tokenisation est la mesure initiale du prétraitement du texte. C'est une tâche critique pour la compréhension des requêtes et des documents. Dans ce processus, le texte non structuré est segmenté en mots discrets et ces mots sont nos unités de traitement. Nous appelons ces mots "jetons". La Tokenisation convertit une chaîne de caractères en une séquence de jetons.
Ces jetons sont souvent appelés termes ou mots, mais il est parfois important de faire une distinction type/jeton. Un jeton est une instance d'une séquence de caractères dans un document particulier qui sont regroupés en une unité sémantique utile pour le traitement. Un type est la classe de tous les jetons contenant la même séquence de caractères. Un terme est un type (peut-être normalisé) inclus dans le dictionnaire du système RI. L'ensemble des termes d'index peut être entièrement distinct des jetons, par exemple, ils peuvent être des identifiants sémantiques dans une taxonomie, mais en pratique dans les systèmes RI modernes, ils sont fortement liés aux jetons dans le document \cite{allahyari2017brief}.\bigskip

Comme pour tous les aspects de la compréhension des requêtes, la tokenisation représente un ensemble de compromis. Une segmentation très littérale de la requête est susceptible d'être bonne pour la précision, mais mauvaise pour le rappel; tandis qu'une approche plus agressive utilisant des tokenisations multiples améliorera le rappel au détriment de la précision. La gestion de cas particuliers tels que les chaînes qui combinent des lettres et des chiffres introduit de la complexité.\bigskip

Pour notre cas on a défini ces limites par les espaces blancs et les caractères de ponctuation. Ces jetons sont maintenant l'entrée d'un autre processus qui est le filtrage des mots vides.

\subsection{Filtrage des mots vides}
Parfois, certains mots extrêmement communs qui sembleraient avoir peu de valeur pour aider à sélectionner des documents correspondant à un besoin de l'utilisateur sont exclus du vocabulaire entièrement. Ces mots sont appelés mots d'arrêt ou  mots vides. La stratégie générale pour déterminer une liste d'arrêt est de trier les termes par fréquence de collecte (le nombre total de fois que chaque terme apparaît dans la collection de documents), puis de prendre les termes les plus fréquents, souvent filtrés manuellement pour leur contenu sémantique.\bigskip

Un exemple de liste d'arrêt est illustré par la figure 3.3. L'utilisation d'une liste d'arrêt réduit considérablement le nombre d'écritures qu'un système doit stocker.

\begin{figure}[H]
\centering 
\resizebox{\linewidth}{!}{\includegraphics[]{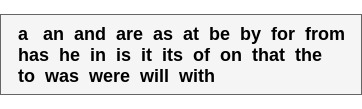}}
\captionof{figure}{Exemple de liste d'arrêt}
\end{figure}

Par conséquent, les filtres sont appliqués aux données et le filtre des mots d'arrêt est une méthode de filtrage standard qui est appliquée en tant que deuxième étape de la phase de prétraitement.
\subsection{Racinisation}
Les différentes formes d'un mot communiquent souvent essentiellement le même sens. Par exemple, il n'y a probablement aucune différence d'intention entre une recherche de "shoe" et une recherche de "shoes".\bigskip

Pour des raisons grammaticales, les documents vont utiliser différentes formes de mots, comme "organize", "organizes" et "organizing". Ces différences syntaxiques entre les formes de mots sont appelées des inflexions, et elles créent des défis pour la compréhension des requêtes. Dans l'exemple de "shoe" et "shoes", il est souhaitable de traiter les deux formes de manière identique mais il n'est pas de même pour les mots "logistic" et "logistics", ce qui signifie des choses très différentes malgré leur similitude apparente. Dans de nombreuses situations, il semble qu'il serait utile que la recherche de l'un de ces mots renvoie des documents contenant un autre mot de l'ensemble.\bigskip

Il existe différents algorithmes qui peuvent être utilisés dans le processus de racinisation.
L'approche de racinisation la plus connue et la plus populaire pour l'anglais est l'algorithme de racinisation de Porter, également connu sous le nom de Porter stemmer, qui s'est avéré empiriquement très efficace à plusieurs reprises. C'est une collection de règles (ou d'heuristiques) conçue pour refléter la façon dont l'anglais gère les inflexions. Par exemple, le stemmer de Porter hache "apple" et "apples" à "appl", et il hache "beyy" et "berries" à "berri".\bigskip

L'algorithme de Porter se compose de 5 phases de réductions de mots distinctes, appliquées séquentiellement les unes après les autres. Dans chaque phase, il existe diverses conventions pour sélectionner des règles, telles que la sélection de la règle de chaque groupe de règles qui s'applique au suffixe le plus long \cite{willett2006porter}.\bigskip

Lors de cette étape nous avons utilisé l'algorithme de Porter. Cette étape nous a permis de réduire chaque mot à sa forme racine et donc on va pouvoir mappé les mots liés à la même racine, même si cette racine n'est pas en elle-même une racine valide.
\section{Phase de création des vecteurs}
Une fois le prétraitement terminé on applique le compteur de fréquence de mot. Lors de cette étape nous avons calculé le nombre d’occurrence de chaque mot de notre jeu données. Après ce calcul nous avons choisit les $n$ mots les plus fréquents. Les mots les plus fréquents sont considérés comme des mots forts ou confiants. Après avoir choisit ces $n$ mots les plus fréquents nous avons procédé au calcul du TF-IDF pour évaluer l'importance de chaque mot de ces $n$ mots les plus fréquents dans chaque document. Nous avons construit un fichier contenant dans chaque ligne le vecteur représentant un document des $m$ documents de notre jeu de données. C'est-à-dire que chaque document est représenté par un ensemble de fonctions de mots clés. Nous traitons chaque document comme un vecteur où les mots de ce document sont considérés comme des caractéristiques ou l'élément de ce vecteur. Le tableau 3.1 montre un exemple de notre fichier. Ce fichier sera l'entrée de la troisième phase.

\begin{table}[H]
\centering
\caption{Vecteurs des documents du jeu de données}
\label{my-label}
\resizebox{\linewidth}{!}{
\begin{tabular}{p{1cm}lllllll}
\cline{1-4} \cline{6-8}
\multicolumn{1}{|l|}{\diagbox{\textbf{Documents}}{\textbf{Mots-clés}}}               & \multicolumn{1}{p{1cm}|}{\textbf{nation}} & \multicolumn{1}{p{1cm}|}{\textbf{state}} & \multicolumn{1}{p{1.5cm}|}{\textbf{graphic}} & \multicolumn{1}{p{1cm}|}{........} & \multicolumn{1}{p{1.5cm}|}{\textbf{machine}} & \multicolumn{1}{p{1cm}|}{\textbf{gun}} & \multicolumn{1}{p{1cm}|}{\textbf{church}} \\ \cline{1-4} \cline{6-8} 
\multicolumn{1}{|l|}{\textbf{20540}} & \multicolumn{1}{p{1cm}|}{0.19}            & \multicolumn{1}{p{1cm}|}{0.284}          & \multicolumn{1}{p{1.5cm}|}{0}                & \multicolumn{1}{p{1cm}|}{........} & \multicolumn{1}{p{1.5cm}|}{0}                & \multicolumn{1}{p{1cm}|}{0}            & \multicolumn{1}{p{1cm}|}{0.675}           \\ \cline{1-4} \cline{6-8} 
........                        & ........                         & ........ & ........                         & ........                      & ........                        & ........                     & ........                         \\ \cline{1-4} \cline{6-8} 
\multicolumn{1}{|l|}{\textbf{54906}} & \multicolumn{1}{p{1cm}|}{0}               & \multicolumn{1}{p{1cm}|}{0.1}            & \multicolumn{1}{p{1.5cm}|}{0}                & \multicolumn{1}{p{1cm}|}{........} & \multicolumn{1}{p{1.5cm}|}{0}                & \multicolumn{1}{p{1cm}|}{0.359}        & \multicolumn{1}{p{1cm}|}{0}               \\ \cline{1-4} \cline{6-8} 
........                        & ........                         & ........                        & ........                         & ........                      & ........                        & ........                     & ........                         \\ \cline{1-4} \cline{6-8} 
\multicolumn{1}{|l|}{\textbf{38543}} & \multicolumn{1}{p{1cm}|}{0}               & \multicolumn{1}{p{1cm}|}{0}              & \multicolumn{1}{p{1.5cm}|}{0.48}             & \multicolumn{1}{p{1cm}|}{........} & \multicolumn{1}{p{1.5cm}|}{0.21}             & \multicolumn{1}{p{1cm}|}{0}            & \multicolumn{1}{p{1cm}|}{0}               \\ \cline{1-4} \cline{6-8} 
........                        & ........                         & ........                        & ........                        & ........                      & ........ & ........                     & ........                        
\end{tabular}}
\end{table}

\section{Phase de clustering}
\subsection{Phase d'apprentissage}
Au cours de cette phase, nous avons utilisé les résultats de la phase précédente, qui est un ensemble de vecteurs représentants chacun un document de notre jeu de données, comme une entrée pour l'algorithme FuzzyART qui est un réseau d'apprentissage par compétition.\medskip

En effet, dans un apprentissage par compétition, rien ne garantit que les catégories formées aillent rester stables. La seule possibilité, pour assurer la stabilité, serait que le coefficient d'apprentissage tende vers zéro, mais le réseau perdrait alors sa plasticité.\smallskip

Les réseaux ART ont été conçus pour harmoniser le dilemme entre l’adaptation (plasticité synaptique) et la stabilité (rigidité synaptique) de l’apprentissage. Dans un système trop plastique, les informations sont apprises même si elles ne sont pas pertinentes; ou les informations apprises sont oubliées. Par contre, dans un système trop rigide, les informations pertinentes ne sont pas apprises.\smallskip

Dans le modèle ART, l’introduction d’un paramètre de vigilance permet de gérer de manière autonome le passage d’un mode plastique à un mode rigide. Les vecteurs de poids ne seront adaptés que si l'entrée fournie est suffisamment proche d'un prototype déjà connu par le réseau. On parlera alors de résonance. Á l'inverse, si l'entrée s'éloigne trop des prototypes existants, une nouvelle catégorie va alors être créée, avec pour prototype, l'entrée qui a engendré sa création.\medskip

Fuzzy ART est un réseau de groupage de données qui est capable d’apprendre à reconnaître des catégories stables, à partir des données analogiques (données réelles comprises entre 0 et 1) ou binaires, ordonnées arbitrairement, pour ce faire, Fuzzy ART utilise des opérateurs de la logique floue.\medskip

FuzzyART va chercher à assigner chaque document c'est-à-dire chaque vecteur de document à une catégorie. Il va nous rendre comme sortie un fichier contenant une liste de chaque cluster avec les documents qui lui appartiennent.\smallskip

La figure 3.4 représente la démarche de la phase d'apprentissage expliquée ci-dessus.
\begin{center}
\begin{figure}[H]
\centering 
\resizebox{\linewidth}{!}{\includegraphics[]{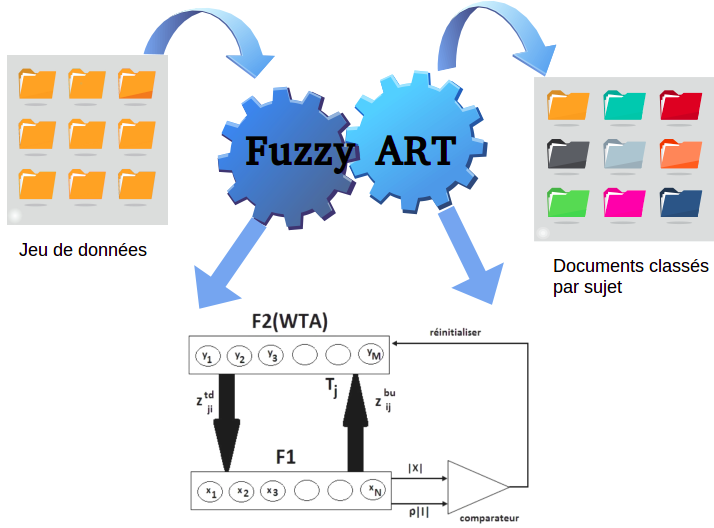}}
\captionof{figure}{Étapes de la phase d'apprentissage}
\end{figure}
\end{center}

Dans une dernière étape nous avons utilisé un classifieur $Paragraph ~Vector$; que nous allons expliquer par la suite; pour la phase de test.
\subsection{Phase de test}
\subsubsection*{Word2Vec}
Word2vec\footnote{\label{ulm}https://deeplearning4j.org/docs/latest/deeplearning4j-nlp-word2vec} est un réseau neuronal à deux couches qui traite le texte. Son entrée est un corpus de texte et sa sortie est un ensemble de vecteurs: vecteurs de caractéristiques pour les mots de ce corpus. Bien que Word2vec ne soit pas un réseau neuronal profond, il transforme le texte en une forme numérique que les réseaux profonds peuvent comprendre.\bigskip

Les applications de Word2vec s'étendent au-delà des phrases d'analyse dans la nature. Il peut être appliqué aussi bien aux gènes, au code, aux goûts, aux playlists, aux graphiques de médias sociaux et à d'autres séries verbales ou symboliques dans lesquelles des modèles peuvent être discernés.\bigskip

Le but et l'utilité de Word2vec est de regrouper les vecteurs de mots similaires dans un espace vectoriel. Autrement dit, il détecte les similitudes mathématiquement. Word2vec crée des vecteurs qui sont des représentations numériques distribuées d'entités de mots, des caractéristiques telles que le contexte de mots individuels. Il le fait sans intervention humaine.\bigskip

Avec suffisamment de données, d'usages et de contextes, Word2vec peut faire des suppositions très précises sur la signification d'un mot en fonction des apparences passées. Ces suppositions peuvent être utilisées pour établir l'association d'un mot avec d'autres mots (par exemple, «man» est «boy», «woman» est «girl»), ou regrouper des documents et les classer par sujet. Ces clusters peuvent constituer la base de la recherche, de l'analyse des sentiments et des recommandations dans des domaines aussi variés que la recherche scientifique, la découverte juridique, le commerce électronique et la gestion de la relation client.\bigskip

La sortie du réseau neuronal Word2vec est un vocabulaire dans lequel chaque élément est associé à un vecteur, qui peut être introduit dans un réseau d'apprentissage en profondeur ou simplement interrogé pour détecter les relations entre les mots.\bigskip

En mesurant la similarité par cosinus, aucune similarité n'est exprimée sous la forme d'un angle de $90$ degrés, alors que la similarité totale de $1$ est un angle de $0$ degré, un chevauchement complet; c'est-à-dire que Sweden est Sweden, tandis que Norway a une distance de $0,760124$ cosinus de Sweden, la plus élevée de tous les autres pays.\bigskip

La figure 3.5 représente une liste de mots associés à "Sweden" en utilisant Word2vec, par ordre de proximité:\medskip

\begin{figure}[H]
 \centering
 \includegraphics[width=8.97cm, height=5.48cm]{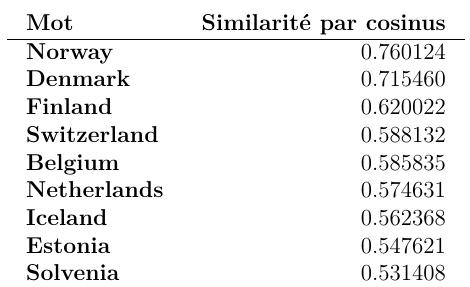}
 \caption{{Mots associés à Sweden avec Word2Vec. } }
 \end{figure}

Un cadre bien connu pour l'apprentissage de Word2Vec est montré dans la figure 3.6. La tâche consiste à prédire un mot étant donné les autres mots dans un contexte.
\bigskip
\bigskip

\begin{figure}[H]
\centering 
\resizebox{\linewidth}{!}{\includegraphics[]{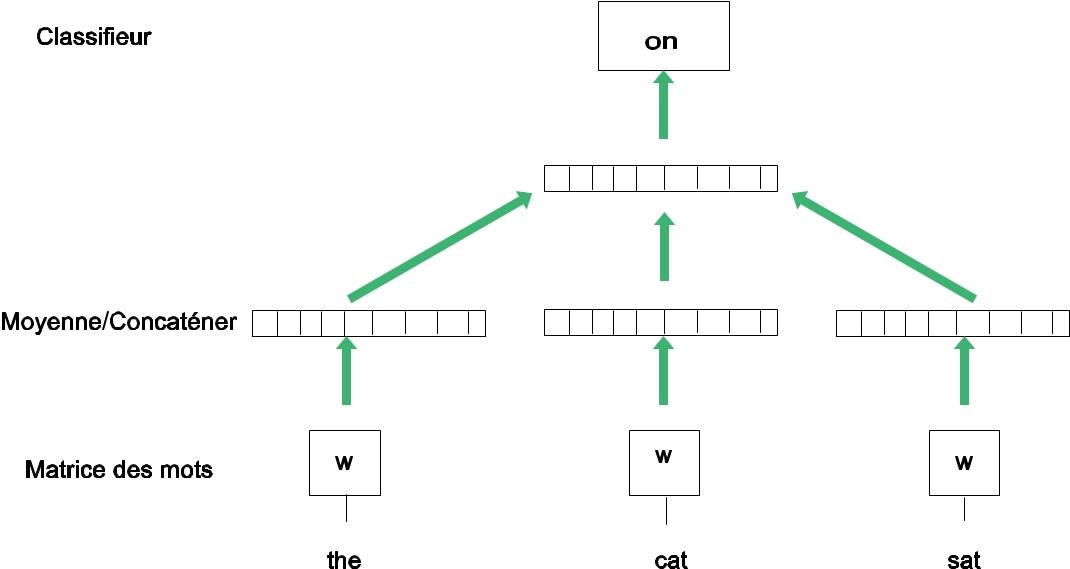}}
\caption{Structure d'apprentissage de Word2Vec \cite{le2014distributed}}
\end{figure} 
\bigskip
\medskip
\bigskip
De façon plus formelle, compte tenu de la séquence des mots de l'apprentissage $w_{1}$, $w_{2}$, $w_{3}$, ..., $w_{T}$, l'objectif du modèle de Word2Vec est de maximiser la probabilité de log moyenne.\medskip
\begin{center}
$\frac{1}{T}\sum_{t=k}^{T-k}\log p(w_{t}|w_{t-k},...,w_{t+k})$
\end{center}

La tâche de prédiction est généralement effectuée via un classificateur multi-classes, tel que softmax. Là, nous avons:\medskip
\begin{center}
$p(w_{t}|w_{t-k},...,w_{t+k})= \frac{{\exp^{y_{w_{t}}}}}{{\sum_{i}\exp^{y_{i}}}}$
\end{center}

Chacun des $y_{i}$ est une probabilité logarithmique non normalisée pour chaque mot de sortie $i$, calculé comme suit:\bigskip

\begin{equation}
y=b+Uh(w_{t-k}, ...,w_{t+k};W)
\end{equation}
où $U$, $b$ sont les paramètres softmax. $h$ est construit par une concaténation ou par une moyenne des vecteurs des mots extraits de $W$.
\subsubsection*{Plongements de mots neuraux}
Les vecteurs que nous utilisons pour représenter les mots sont appelés plongements de mots neuraux (neural word embeddings). Une chose en décrit une autre, même si ces deux choses sont radicalement différentes.\bigskip

Word2vec est similaire à un autoencoder, encodant chaque mot dans un vecteur, mais plutôt que de s'entraîner contre les mots d'entrée par reconstruction, comme le fait une machine de Boltzmann restreinte, word2vec entraîne des mots contre d'autres mots qui les voisent dans le corpus d'entrée.\bigskip

Il le fait de deux façons, soit en utilisant un contexte pour prédire un mot cible (une méthode connue sous le nom de sac continu de mots, ou CBOW), soit en utilisant un mot pour prédire un contexte cible, appelé skip-gram. Cette dernière produit des résultats plus précis sur de grands ensembles de données.\bigskip

\begin{figure}[H]

  \begin{minipage}[CBOW]{0.5\linewidth}
   \centering
   \includegraphics[scale=0.5]{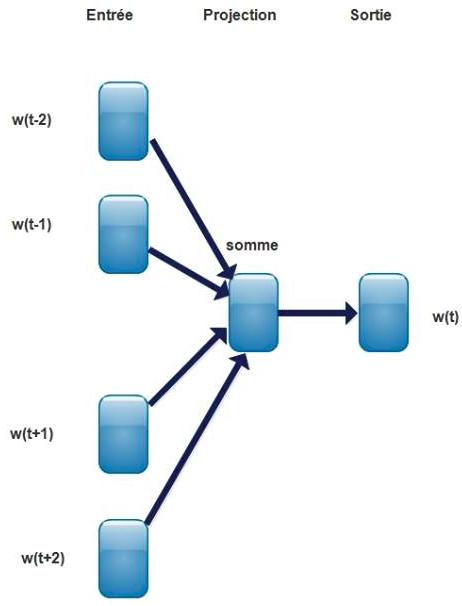} 
   \subcaption{CBOW}     
  \end{minipage} \hfill
  \begin{minipage}[Skip-grap]{0.5\linewidth}
   \centering
   \includegraphics[scale=0.5]{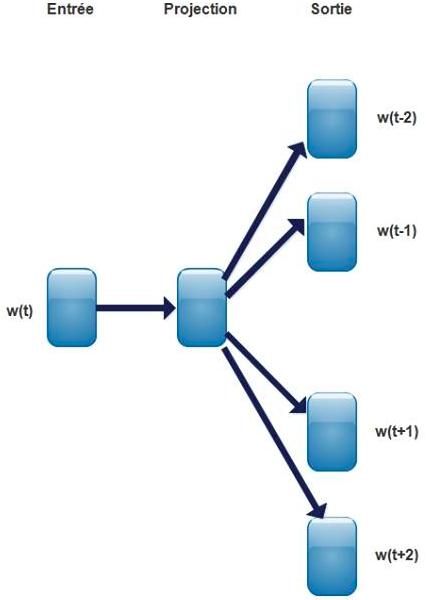} 
   \subcaption{Skip-gram}     
  \end{minipage}
  \label{fig:ma_fig}

\end{figure}

Lorsque le vecteur de caractéristique assigné à un mot ne peut pas être utilisé pour prédire avec précision le contexte de ce mot, les composantes du vecteur sont ajustées. Le contexte de chaque mot dans le corpus est l'enseignant envoyant des signaux d'erreur en retour pour ajuster le vecteur caractéristique. Les vecteurs de mots jugés similaires par leur contexte sont rapprochés en ajustant les nombres dans le vecteur.\bigskip

Un ensemble de vecteurs de mots qui a fait un bon apprentissage, placera des mots similaires proches les uns des autres dans un espace vectoriel à n dimensions. Les mots "oak", "elm" et "birch" peuvent se regrouper dans un coin, tandis que "war", "conflict" et "strife" se regrouperaient dans un autre.\bigskip

Des choses et des idées similaires sont présentées comme étant "proches". Leurs significations relatives ont été traduites en distances mesurables. Les qualités deviennent des quantités, et les algorithmes peuvent faire leur travail. Mais la similarité n'est que la base de nombreuses associations que Word2vec peut apprendre. Par exemple, il peut mesurer les relations entre les mots d'une langue et les relier à une autre.\bigskip

Ces vecteurs sont la base d'une géométrie plus complète des mots. Non seulement Rome, Paris, Berlin et Beijing se rapprocheront l'une de l'autre, mais elles auront chacune des distances similaires en espace vectoriel avec les pays dont elles sont les capitales; c'est-à-dire Rome - Italy = Beijing - China. Et si on sait seulement que Rome était la capitale de Italy et qu'on se pose des questions sur la capitale de China, alors l'équation "Rome - Italy + China" reviendrait à Beijing.\bigskip 

\begin{figure}[H]
\centering 
\resizebox{0.93\linewidth}{!}{\includegraphics[]{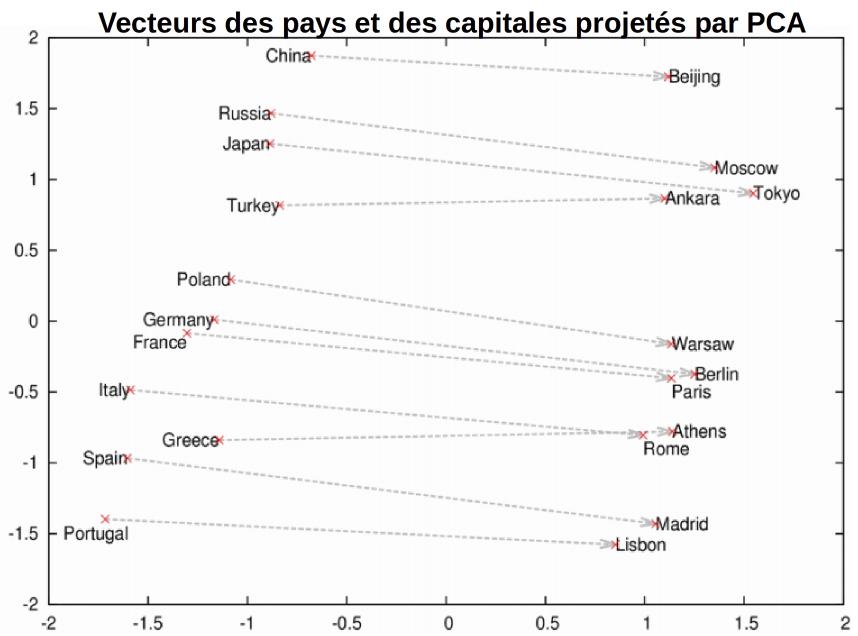}}
\captionof{figure}{Projection des vecteurs des pays et des capitales par PCA}
\end{figure}

\subsubsection*{Présentation de paragraph vector}
Paragraph Vector \cite{le2014distributed}, est un algorithme non supervisé qui apprend des représentations d'entités de longueur fixe à partir de morceaux de textes de longueur variable, tels que des phrases, des paragraphes et des documents.\bigskip

C'est une technique inspirée par le travail récent d'apprentissage de représentations vectorielles de mots utilisant des réseaux de neurones (word2vec).\medskip

Paragraph Vector est capable de construire des représentations de séquences d'entrée de longueur variable. Contrairement à certaines approches précédentes, il est général et applicable aux textes de toute longueur: phrases, paragraphes et documents. Il ne nécessite pas de réglage spécifique à la tâche de la fonction de pondération des mots et ne dépend pas non plus des arbres d'analyse. \bigskip

Dans paragraph vector, la structure de la softmax hiérarchique est un arbre Huffman binaire, où les codes courts sont affectés à des mots fréquents.\\
\subsubsection*{Paragraph Vector: Un modèle de mémoire distribuée}
L'approche pour l'apprentissage de paragraph vector \cite{le2014distributed} est inspirée par les méthodes d'apprentissage de Word2Vec. L'inspiration est que Word2Vec est invité à contribuer à une tâche de prédiction sur le mot suivant dans la phrase. Ainsi, malgré le fait que les vecteurs de mots sont initialisés de manière aléatoire, ils peuvent éventuellement capturer la sémantique en tant que résultat indirect de la tâche de prédiction. Cette idée sera utilisée dans paragraph vector d'une manière similaire. Paragraph vector est également invité à contribuer à la tâche de prédiction du mot suivant compte tenu des nombreux contextes échantillonnés à partir du paragraphe.\bigskip

Dans Paragraph Vector (voir Figure 3.9), chaque paragraphe est mappé à un vecteur unique, représenté par une colonne dans la matrice $D$ et chaque mot est également mappé à un vecteur unique, représenté par une colonne dans la matrice $W$. Paragraph Vector et Word2Vec sont moyennés ou concaténés pour prédire le mot suivant dans un contexte.\medskip

Plus formellement, le seul changement dans ce modèle par rapport au cadre de Word2Vec est dans l'équation 3.1, où $h$ est construit à partir de $W$ et $D$.\bigskip

Le jeton de paragraphe peut être considéré comme un autre mot. Il agit comme une mémoire qui se souvient de ce qui manque dans le contexte actuel ou le sujet du paragraphe. Pour cette raison, ce modèle est souvent appelé le modèle de mémoire distribuée de Paragraph Vector (PV-DM).\bigskip

Paragraph Vector et Word2Vec sont formés en utilisant la descente de gradient stochastique et le gradient est obtenu par rétro-propagation. À chaque étape de la descente de gradient stochastique, on peut échantillonner un contexte de longueur fixe à partir d'un paragraphe aléatoire, calculer le gradient d'erreur du réseau de la figure 3.8 et utiliser le gradient pour mettre à jour les paramètres du modèle.\bigskip

Au moment de la prédiction, il est nécessaire d'effectuer une étape d'inférence pour calculer le vecteur de paragraphe pour un nouveau paragraphe. Ceci est également obtenu par descente du gradient. Dans cette étape, les paramètres pour le reste du modèle, les vecteurs de mots $W$ et les poids softmax, sont fixés. Supposons qu'il y ait $N$ paragraphes dans le corpus, $M$ mots dans le vocabulaire, et que nous voulions apprendre les vecteurs de paragraphes de telle sorte que chaque paragraphe soit mappé à $p$ dimensions et que chaque mot soit mappé à $q$ dimensions, alors le modèle a le total de $ N \times p + M \times q$ paramètres (à l'exclusion des paramètres softmax). Même si le nombre de paramètres peut être important lorsque $N$ est grand, les mises à jour pendant l'apprentissage sont généralement rares et donc efficaces.\bigskip

\begin{figure}[H]
\centering 
\resizebox{\linewidth}{!}{\includegraphics[]{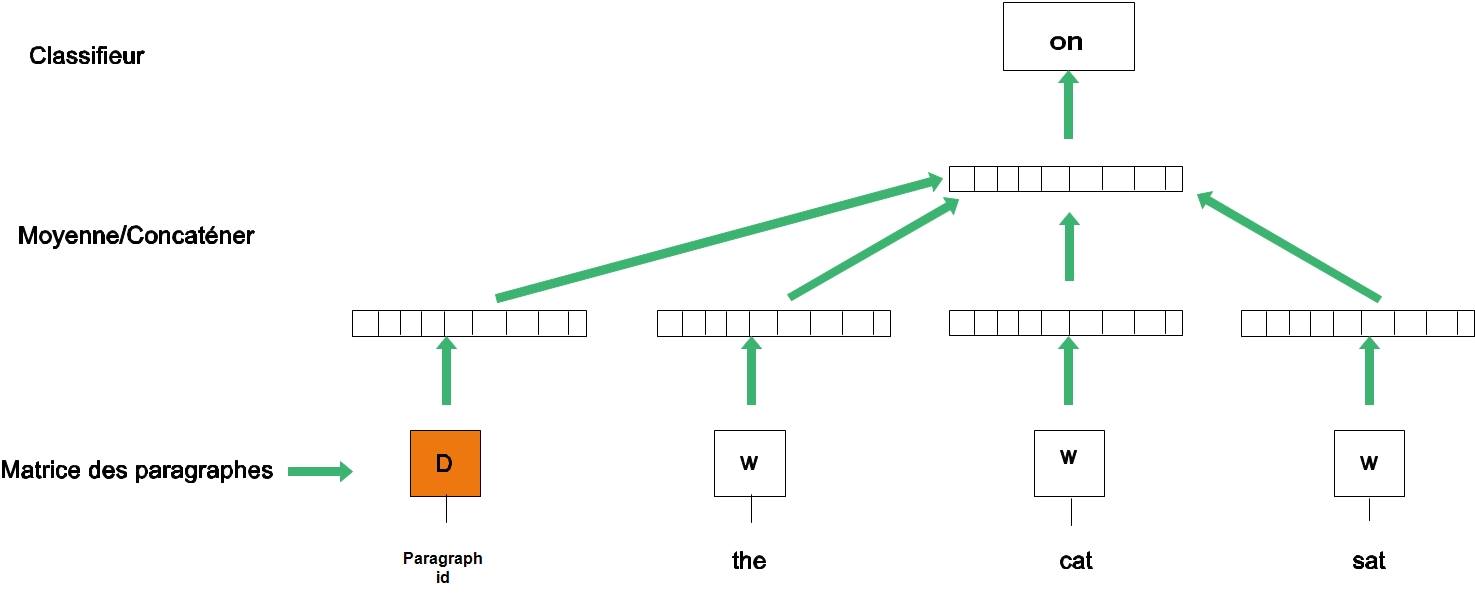}}
\captionof{figure}{Modèle de mémoire distribuée \cite{le2014distributed}}
\end{figure}
Après l'apprentissage, les vecteurs de paragraphe peuvent être utilisés en tant que caractéristiques pour le paragraphe. Nous pouvons appliquer ces caractéristiques directement aux techniques d'apprentissage automatique conventionnelles telles que la régression logistique, les machines vectorielles de support ou les K-means.\bigskip

En résumé, l'algorithme lui-même comporte deux étapes clés: 
\begin{itemize}
\item L'apprentissage non supervisé pour obtenir des vecteurs de mot $W$, des poids softmax $U$, $b$ et des vecteurs de paragraphe $D$ sur des paragraphes déjà vus.
\item L'étape d'inférence pour obtenir les vecteurs de paragraphe $D$ pour les nouveaux paragraphes (jamais vu auparavant) en ajoutant plus de colonnes dans $D$ et le gradient descendant sur $D$ en maintenant $W$, $U$, $b$ fixe.
\end{itemize}
\subsubsection*{Paragraph Vector sans ordre de mots: Sac de mots  distribué}
La méthode ci-dessus \cite{le2014distributed} considère la concaténation de Paragraph vector avec Word2Vec pour prédire le mot suivant dans un texte. Une autre méthode consiste à ignorer les mots de contexte dans l'entrée, mais à forcer le modèle à prédire des mots échantillonnés aléatoirement à partir du paragraphe de la sortie. En réalité, cela signifie qu'à chaque itération de la descente de gradient stochastique, nous échantillonnons une fenêtre de texte, puis échantillonnons un mot aléatoire de la fenêtre de texte et formons une tâche de classification en fonction de paragraph vector. Cette technique est illustrée par la figure 3.10. Cette version est nommée la version distribuée du sac de mots de paragraph Vector (PV-DBOW), par opposition à la version de mémoire distribuée de Paragraph Vector (PV-DM) dans la section précédente.\bigskip

\begin{figure}[H]
\centering 
\resizebox{\linewidth}{!}{\includegraphics[]{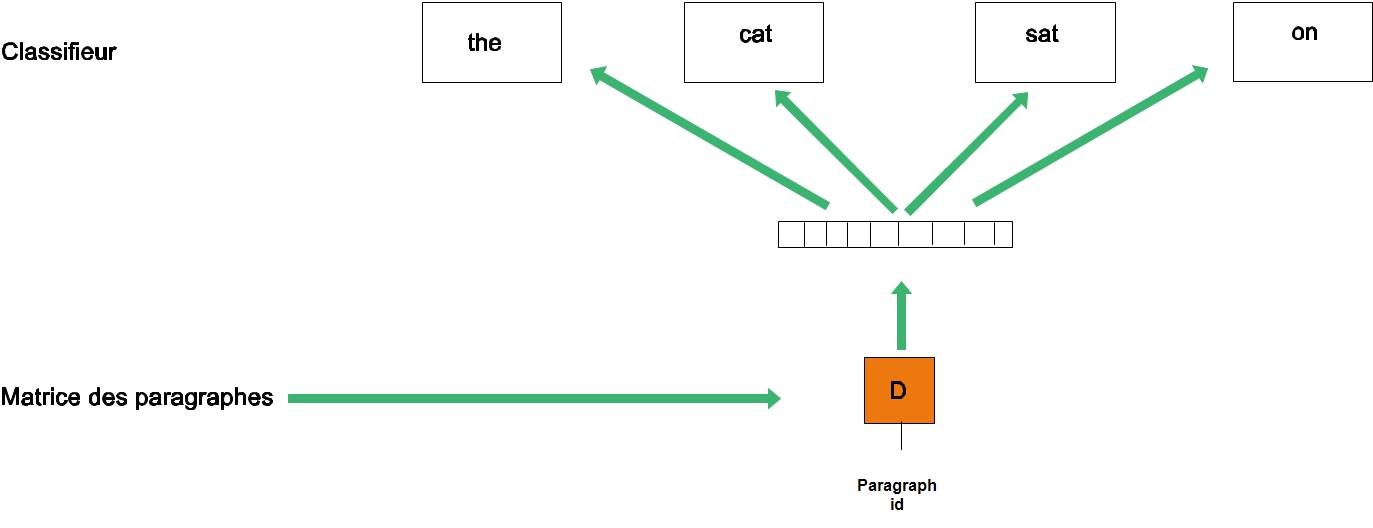}}
\captionof{figure}{Modèle de sac de mots  distribué \cite{le2014distributed}}
\end{figure}

En plus d'être conceptuellement simple, ce modèle nécessite de stocker moins de données. Nous avons seulement besoin de stocker les poids softmax par opposition aux poids softmax et aux vecteurs de mots dans le modèle précédent. Ce modèle est également similaire au modèle de Skip-gram dans Word2Vec.
\subsubsection*{Les avantages de paragraph vector:}
Un avantage important de Paragraph Vector est qu'il est obtenu à partir de données non étiquetées et peut donc bien fonctionner pour des tâches qui n'ont pas suffisamment de données étiquetées.\medskip

Paragraph vector traite également certaines des principales faiblesses des modèles de sac de mots. Tout d'abord, il hérite d'une propriété importante de Word2Vec: la sémantique des mots. Dans cet espace, "powerful" est plus proche de "strong" que de "Paris". Le deuxième avantage de paragraph vector est qu'il prend en considération l'ordre des mots, au moins dans un petit contexte, de la même manière qu'un modèle n-gram avec un $n$ grand le ferait. Ceci est important, car le modèle n-gram conserve beaucoup d'informations sur le paragraphe, y compris l'ordre des mots.
\subsection*{L'utilisation de Paragraph Vector pour la phase de test}
Après l'obtention d'un ensemble de sujets en provenance de FuzzyART et pour rester dans la thématique des réseaux de neurones on a pensé à utiliser un classifieur utilisant Paragraph Vector pour la phase de test. Le classifieur Paragraph Vector procède comme suit:\medskip

\begin{itemize}
\item Tout d'abord il prend en entrée la liste des sujets associés avec les documents qui leur appartiennent et cette liste provient du résultat que nous a donné FuzzyART. Il prend aussi un ensemble de données de test.
\item Dans un deuxième temps il va tester ces données de test qui ne sont pas étiquetées, pour donner leur similitude à chaque sujet des sujets que FuzzyART a définit.
\end{itemize}
La figure 3.11 est un schéma illustratif qui représente les étapes de cette dernière phase.
\begin{center}
\begin{figure}[H]
\centering 
\resizebox{\linewidth}{!}{\includegraphics[]{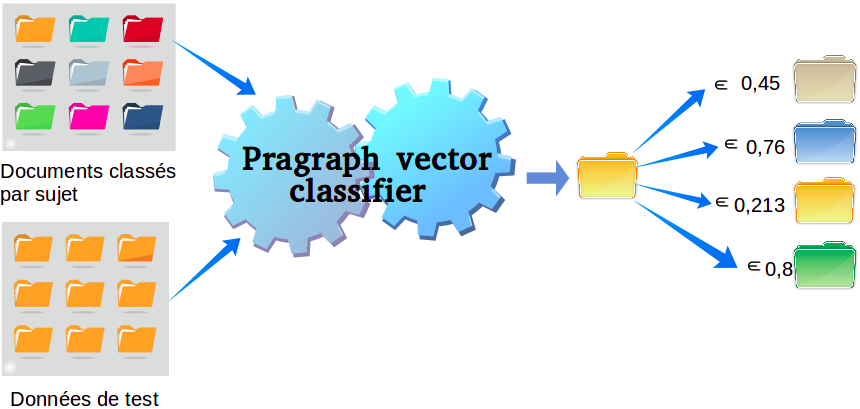}}
\captionof{figure}{Étapes de la phase de test}
\end{figure}
\end{center}

\section{Pseudo algorithme de ClusART}
\begin{algorithm}[H]
{
\LinesNumbered
\Entree{D = \{${d}_1$, ${d}_2$,..., ${d}_n$\}: Ensemble des documents \`a classifier;}
 \Sortie{topic = \{${t}_1$, ${t}_2$,..., ${t}_n$\} : Ensemble de clusters \'etiquettes de D;}
\Deb
{
\PourCh{$d$ $\in$ $D$}{
\PourCh{$mot$ $\in$ $d$}{
\Si{$mot$ $\in$ stopwords}{
\'eliminer($mot$, $d$)};\\
\Sinon{
$mot^{ '}$ = PorterStemmer(mot) ;\\}}}
\PourCh{$mot^{ '}$ $\in$ $D$}{
$tabOcc$ = WordCount($mot^{ '}$) ;}
motRec = r\'eccurent(n, tabOcc);\\
\PourCh{$d$ $\in$ $D$}{
dVec = vecCalc(d, motRec);}
\PourCh{$dVec$}{
r\'esultatFA = FuzzyART(dVec);}
topic = ParagraphVectorClassifier(resultatFA, testSet);

}
}

  \caption{ClusART}
  \label{unoalgo}
\end{algorithm}
\bigskip

\bigskip

On choisit de faire le traitement qui va suivre sur chaque document "\textbf{d}" de notre jeu de données "\textbf{D}". Par la suite on va choisir chaque "\textbf{mot}" de notre document "\textbf{d}" (ligne 3). Si ce "\textbf{mot}" appartient à la liste des mots vides on va l'éliminer en utilisant la fonction \textbf{éliminer} qui prend en paramètres le "\textbf{mot}" et le document "\textbf{d}" contenant ce "\textbf{mot}" (ligne 5). \smallskip

Sinon si "\textbf{mot}" n'appartient pas à la liste des mots vides on va le remplacer par sa racine en utilisant l'algorithme de PorterStemmer. On va donc remplacer "\textbf{mot}" par "\textbf{$mot^{ '}$}" dans "\textbf{d}" (lignes 8).\medskip

On va parcourir tout notre jeu de données "\textbf{$mot^{ '}$}" par "\textbf{$mot^{ '}$}" pour calculer le nombre d’occurrence de chaque "\textbf{mot}" dans le jeu de données en utilisant la fonction \textbf{récurrent} qui prend en paramètres un nombre "\textbf{n}" et le tableau "\textbf{tabOcc}". Le résultat de cette étape sera stocké dans un tableau "\textbf{tabOcc}" contenant comme clé "\textbf{mot}" et comme valeur le nombre d’occurrence de "\textbf{mot}" dans "\textbf{D}" (ligne 10).\medskip

On va par la suite parcourir "\textbf{tabOcc}" pour en extraire les \textbf{n} mots les plus récurrents et on va les stocker dans le tableau "\textbf{motRec}" (ligne 11).\medskip

On va calculer pour chaque document "\textbf{d}" de notre jeu de données "\textbf{D}" son vecteur "\textbf{dVec}" avec comme caractéristiques les mots stockés dans le tableau "\textbf{motRec}". Pour ce faire on va utiliser la fonction \textbf{vecCalc} qui prend en paramètres le document "\textbf{d}" et le tableau "\textbf{motRec}" (ligne 13).\medskip

On va utiliser \textbf{fuzzyART}, qui aura comme entrée l'ensemble des vecteurs "\textbf{dVec}", pour trouver l'ensemble des clusters qui sont les sujets des documents "\textbf{d}" de notre jeu de données "\textbf{D}". On aura alors comme résultat "\textbf{résultatFA}" qui contiendra comme clé les \textbf{clusters(sujets)} et comme valeur l'ensemble des documents "\textbf{d}" appartenant à ce cluster (ligne 15).\medskip

Dans la dernière phase on va utiliser \textbf{ParagraphVectorClassifier} qui prend comme entrée le résultat de l'étape précédente et un ensemble de données de test "\textbf{testSet}". Lors de cette étape on va trouver les clusters (sujets) des données de test "\textbf{testSet}" suivant l'apprentissage de notre jeu de données qu'on a effectué avec \textbf{FuzzyART} dans l'étape précédente et on va stocker le résultat dans "\textbf{topic}".
\addcontentsline{toc}{section}{Conclusion}
\section*{Conclusion}
Durant le présent chapitre nous avons présenté notre nouvelle approche pour la détection de sujets basée sur les réseaux de neurones adaptatifs ART, intitulée ClusART.
Nous avons ainsi décrit les différentes phases.
Dans le chapitre suivant, nous menons une étude expérimentale à travers laquelle nous allons conduire différentes expérimentations permettant d’évaluer les performances de notre approche pour la détection de sujets.
\chapter{Expérimentation et évaluation}
\addcontentsline{toc}{section}{Introduction}
\section*{Introduction}
Durant ce chapitre nous nous consacrons à l’évaluation de notre approche pour la détection des sujets. Après la présentation de notre environnement d’expérimentations et la description de la collection de test, nous procéderons à une série d’expérimentations visant à promouvoir les performances de notre approche pour la détection de sujets.
\section{Environnement de travail}
L'ensemble de nos expérimentations ont été réalisées sur un PC muni du système d'exploitation ubuntu 14.04 (64bits) avec un processeur Intel® Core™ i5-4200U et une fréquence d'horloge de 1.60GHz. Pour le développement de notre approche nous avons utilisés:
\begin{itemize}
\item \textbf{Java :} C'est un langage de programmation orienté objet créé par James Gosling et Patrick Naughton, employés de Sun Microsystems, avec le soutien de Bill Joy (cofondateur de Sun Microsystems en 1982), présenté officiellement le 23 mai 1995 au SunWorld.

\item \textbf{IntelliJ IDEA :} Le logiciel IntelliJ IDEA est un IDE commercial pour Java développé par JetBrains. Il se distingue par une meilleure vitesse de compilation et de réaction que ses concurrents.\medskip

Les langages de programmations s'intégrant à son IDE sont les suivants : Java, JavaScript, CoffeeScript, HTML, XHTML, CSS, XML, XSL, ActionScript, MXML, Python, Ruby, JRuby, Groovy, SQL, PHP, Scala et Kotlin. De plus, les cadres d'applications et technologies supportés pour l'IDE sont les suivants : Ajax, Android, Apache Tapestry, Django, EJB, FreeMarker, Hibernate/JPA, JBoss Seam, JSP, JSF, Google App Engine, Google Web Toolkit, Grails, Java ME MIDP/CLDC, OSGi, Play, Ruby on Rails, Struts, Struts 2, Spring, Web Services, Velocity.
\end{itemize}

\section{Description des données de test}
Afin d'expérimenter l'approche de recherche proposée, il est nécessaire de disposer d'un jeu de données comportant l'ensemble des variables sélectionnées. En outre ce jeu de données devrait être issu d'un domaine d'application où il existe des ressources présentant des données adaptées à notre contexte. Nous avons ainsi exploité la collection 20 Newsgroups.\medskip

L'ensemble de 20 Newsgroups comprend environ 18000 Newsgroups sur 20 sujets répartis en deux sous-ensembles: un pour l'apprentissage (ou le développement) et l'autre pour les tests (ou pour l'évaluation des performances).\bigskip

Les données sont organisées en 20 groupes de discussion différents, correspondant chacun à un sujet différent. Les sujets de la catégorie sont liés à \textit{computers}, \textit{politics}, \textit{religion}, \textit{sports} et \textit{science}. Chaque document appartient à exactement un groupe de discussion, mais il y a une petite fraction des articles qui appartiennent à plus d'une catégorie. Certains des groupes de discussion sont très proches les uns des autres (par exemple comp.sys.ibm.pc.hardware / comp.sys.mac.hardware), tandis que d'autres sont hautement indépendants (par exemple misc.forsale / soc.religion.christian). Le tableau 4.1 représente la liste des 20 Newsgroups, partitionnés (plus ou moins) en fonction du sujet:\bigskip

\begin{table}[H]
\centering
\caption{Les sujets des 20 Newsgroups}
\begin{tabular}{|l|l|l|}
\hline
\begin{tabular}[c]{@{}l@{}}comp.graphics \\ comp.os.ms-windows.misc\\ comp.sys.ibm.pc.hardware\\ comp.sys.mac.hardware\\ comp.windows.x\end{tabular} & \begin{tabular}[c]{@{}l@{}}rec.autos\\ rec.motorcycles\\ rec.sport.baseball\\ rec.sport.hockey\end{tabular} & \begin{tabular}[c]{@{}l@{}}sci.crypt\\ sci.electronics\\ sci.med\\ sci.space\end{tabular}         \\ \hline
misc.forsale                                                                                                                                         & \begin{tabular}[c]{@{}l@{}}talk.politics.misc\\ talk.politics.guns\\ talk.politics.mideast\end{tabular}     & \begin{tabular}[c]{@{}l@{}}talk.religion.misc\\ alt.atheism\\ soc.religion.christian\end{tabular} \\ \hline
\end{tabular}

\end{table}

\begin{table}[H]
\centering
\caption{Documents dans 20 Newsgroups}
\resizebox{\linewidth}{!}{
\footnotesize{
\begin{tabular}{|l|l|l|l|}
\hline
\textbf{Sujet}                   & \textbf{Document en formation} & \textbf{Document en test} & \textbf{Total document} \\ \hline
\textbf{alt.atheism }             & 480                   & 319              & 799            \\ \hline
\textbf{comp.graphics}            & 584                   & 389              & 973            \\ \hline
\textbf{comp.os.ms-windows.misc}  & 572                   & 394              & 966            \\ \hline
\textbf{comp.sys.ibm.pc.hardware} & 590                   & 392              & 982            \\ \hline
\textbf{comp.sys.mac.hardware}    & 578                   & 385              & 963            \\ \hline
\textbf{comp.windows.x}           & 593                   & 392              & 985            \\ \hline
\textbf{misc.forsale}             & 585                   & 390              & 975            \\ \hline
\textbf{rec.autos}                & 594                   & 395              & 989            \\ \hline
\textbf{rec.motorcycles}          & 598                   & 398              & 996            \\ \hline
\textbf{rec.sport.baseball}       & 597                   & 397              & 994            \\ \hline
\textbf{rec.sport.hockey}         & 600                   & 399              & 999            \\ \hline
\textbf{sci.crypt}                & 595                   & 396              & 991            \\ \hline
\textbf{sci.electronics}          & 591                   & 393              & 984            \\ \hline
\textbf{sci.med}                  & 594                   & 396              & 990            \\ \hline
\textbf{sci.space}               & 593                   & 394              & 987            \\ \hline
\textbf{soc.religion.christian}   & 598                   & 398              & 996            \\ \hline
\textbf{talk.politics.guns}       & 545                   & 364              & 909            \\ \hline
\textbf{talk.politics.mideast}    & 564                   & 376              & 940            \\ \hline
\textbf{talk.politics.misc}       & 465                   & 310              & 775            \\ \hline
\textbf{talk.religion.misc}       & 377                   & 251              & 628            \\ \hline
\textbf{Total}                    & 11293                 & 7528             & 18821          \\ \hline
\end{tabular}}}
\end{table}

\section{Métriques d’évaluation}
Dans l'objectif d'évaluer la performance de notre approche de détection de sujet, l'usage d'un ensemble de métriques est primordiale. Nous avons ainsi utilisé les métriques communément utilisées dans le domaine de la recherche d'information, à savoir:
\begin{enumerate}
\item \textbf{La précision}: Elle représente le nombre de documents pertinents retrouvés par le système parmi la liste de tous les documents retournés. Cette mesure est calculée comme suit:
\smallskip

\begin{center}
\begin{eqnarray*}
\textbf{précision} = \frac{\textbf{nombre de documents pertinents retrouvés}}{\textbf{nombre total de documents retrouvés}}
\end{eqnarray*}
\bigskip
\end{center}

\item \textbf{Le rappel}: Il représente le nombre de documents pertinents retrouvés par le système parmi la liste de tous les documents pertinents existants au sein de ce dernier. Celui-ci est exprimée par:
\smallskip

\begin{centering}
\begin{eqnarray*}
\textbf{rappel} = \frac{\textbf{nombre de documents pertinents retrouvés}}{\textbf{nombre total de documents pertinents dans la collection}}
\end{eqnarray*}
\end{centering}
\bigskip

\item \textbf{La F-mesure}: Elle correspond à l'exactitude d'un test et regroupe la précision et le rappel qui sont deux métriques complémentaires c'est-à-dire quand la valeur de la première augmente , celle du deuxième diminue et vice-versa. Une telle mesure est exprimée par la formule ci-dessous:
\smallskip

\begin{center}
\begin{eqnarray*}
\textbf{${F}_{mesure}$} = \textbf{2 * }\frac{\textbf{précision * rappel}}{\textbf{précision + rappel}}
\end{eqnarray*}
\end{center}

\end{enumerate}

\section{Résultats expérimentaux et discussion}
\subsection*{Étude de l'impact du paramètre de vigilance $\rho$}
Le réseau Fuzzy ART est contrôlé par trois paramètres : le paramètre de sélection, le paramètre d’apprentissage et le paramètre de vigilance.\bigskip

\begin{itemize}
\item Le paramètre de sélection $\alpha$ est utilisé dans le calcul des fonctions de sélection. Lorsque $\alpha$ tend vers 0, le recodage (neurone gagnant différent pour un même vecteur d’entrée) est minimisé lors de l’apprentissage. Aussi, quand $\alpha$ augmente, le réseau a tendance à créer plus de classes. Le phénomène constaté lorsqu’on augmente la valeur du paramètre de sélection $\alpha$ est l’augmentation du nombre de classes créées par le réseau, plus le paramètre est grand, plus il y a prolifération du nombre de classes. Si le réseau crée plus de classes quand $\alpha$ est grand, c’est que la région d’attraction autour des frontières des classes diminue. Par ailleurs, quand le paramètre de sélection devient relativement grand, la vigilance perd de son effet. Il semble donc plus avantageux de garder la valeur du paramètre de sélection $\alpha$ assez petit. Nous avons fixé ce paramètre à 0.2.\bigskip

\item Le paramètre d’apprentissage $\beta$ est utilisé lors de la mise à jour des poids de connexions. Il influence la vitesse à laquelle les poids du réseau sont modifiés lors de l’apprentissage. Plus $\beta$ est petit, plus les poids varient lentement. \'A la limite, si $\beta$ =0, les poids ne varieront jamais. Par contre, si $\beta$ =1, le réseau va avoir un apprentissage rapide. Nous avons attribué à ce paramètre la valeur de 0.4.\bigskip

\item Le paramètre de vigilance $\rho$ sert de critère pour déterminer si la classe (le neurone gagnant) choisie par le réseau est acceptée ou non. Plus $\rho$ est petit, plus les classes créées sont grossières, tandis que plus $\rho$ est grand, plus les classes créées sont précises. Ce paramètre impacte directement le nombre de classes créé par le réseau lors de l’apprentissage. Plusieurs tests nous ont conduits à fixer la valeur de ce paramètre à 0.8. 
\end{itemize}
\bigskip

Le graphe de la figure 4.1 représente la variation du nombre de classes construites par le réseau Fuzzy ART selon les différentes valeurs attribuées au paramètre de vigilance. On constate que le nombre exact des classes est obtenu en attribuant à $\rho$ la valeur 0.8.\bigskip

\begin{figure}[H]
\centering 
\resizebox{\linewidth}{!}{\includegraphics[]{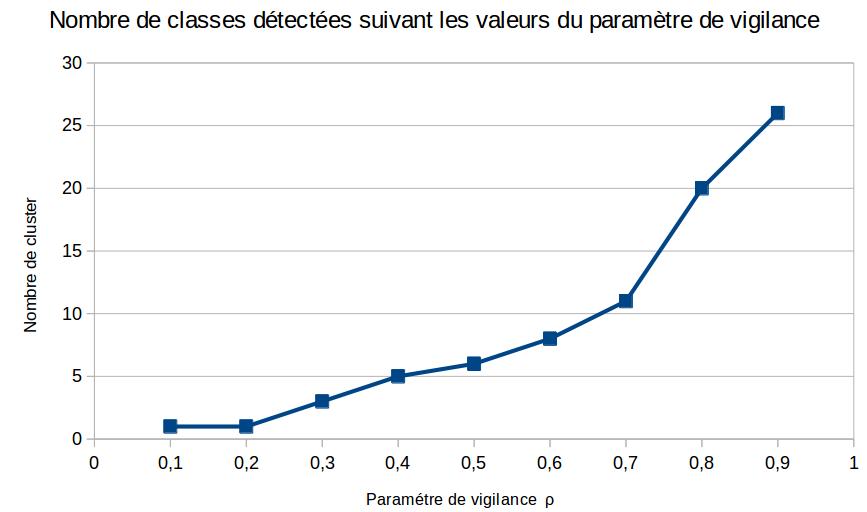}}
\captionof{figure}{Nombre de classes détectées suivant les valeurs du paramètre de vigilance}
\end{figure}

\subsection*{Évaluation de ClusART}

Pour tester la performance de l'approche proposée pour la détection de sujet des documents, nous avons mené une étude comparative avec quelques algorithmes.\smallskip

La détection de sujet est la classification, la différenciation et la détection de, par exemple du texte, dans un seul programme. Des approche communes utilisent LDA et KNN.
Dans cette expérience, nous comparons notre algorithme de détection de sujet avec les deux algorithmes qui sont les suivants :\bigskip

\begin{itemize}
\item KNN : L'algorithme de regroupement k-Nearest Neighbor est une solution de regroupement populaire pour les tâches TDT(topic detection and tracking) [Allan et al 1998].\smallskip

\item LDA : Il s'agit d'un modèle probabiliste génératif pour les collections de données discrètes telles que les corpus de texte. C'est un modèle bayésien hiérarchique à trois niveaux, dans lequel chaque élément d'une collection est modélisé comme un mélange fini sur un ensemble de sujets sous-jacent.[Blei 2003]
\end{itemize}

\subsection*{- Mesures pour 5000 documents}

\begin{table}[H]
\centering
\caption{Valeurs de rappel, précision et F-mesure obtenues pour 5000 documents}
\label{my-label}
\begin{tabular}{l|l|l|l|}
\cline{2-4}
                              & Précision     & Rappel        & F-mesure      \\ \hline
\multicolumn{1}{|l|}{KNN}     & 0,59          & 0,49          & 0,53          \\ \hline
\multicolumn{1}{|l|}{LDA}     & \textbf{0,69} & 0,62          & 0,65          \\ \hline
\multicolumn{1}{|l|}{ClusART} & 0,59          & \textbf{0,76} & \textbf{0,66} \\ \hline
\end{tabular}
\end{table}

\begin{figure}[H]
\centering 
\resizebox{\linewidth}{!}{\includegraphics[]{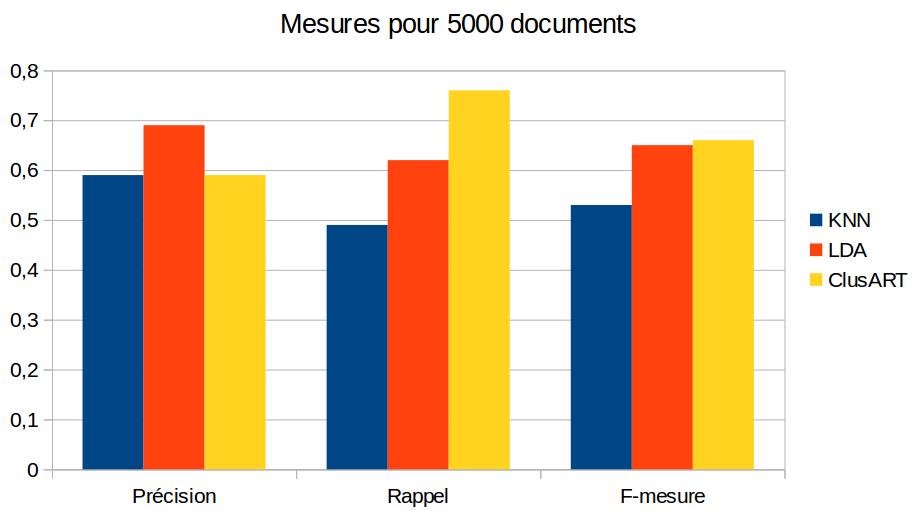}}
\captionof{figure}{Mesure de performance pour 5000 documents}
\end{figure}
\bigskip

La figure 4.2 et le tableau 4.3 rapportent les valeurs de rappel, précision et F-mesure pour 5000 documents. Il est intéressant de noter que, selon les histogrammes esquissés, nous pouvons souligner que ClusART surpasse nettement KNN et LDA. En fait leurs valeurs de rappel sont beaucoup plus faibles que celle obtenue par notre approche. Cependant, la précision de KNN et LDA surpasse légèrement notre approche. La performance de notre approche est confirmée en terme de F-mesure, elle surpasse aussi KNN et LDA.  
\subsection*{- Mesures pour 10000 documents}

\begin{table}[H]
\centering
\caption{Valeurs de rappel, précision et F-mesure obtenues pour 10000 documents}
\label{my-label}
\begin{tabular}{l|l|l|l|}
\cline{2-4}
                              & Précision     & Rappel        & F-mesure     \\ \hline
\multicolumn{1}{|l|}{KNN}     & 0,51          & 0,42          & 0,46         \\ \hline
\multicolumn{1}{|l|}{LDA}     & \textbf{0,53} & 0,45          & 0,48         \\ \hline
\multicolumn{1}{|l|}{ClusART} & 0,39          & \textbf{0,72} & \textbf{0,5} \\ \hline
\end{tabular}
\end{table}

\begin{figure}[H]
\centering 
\resizebox{\linewidth}{!}{\includegraphics[]{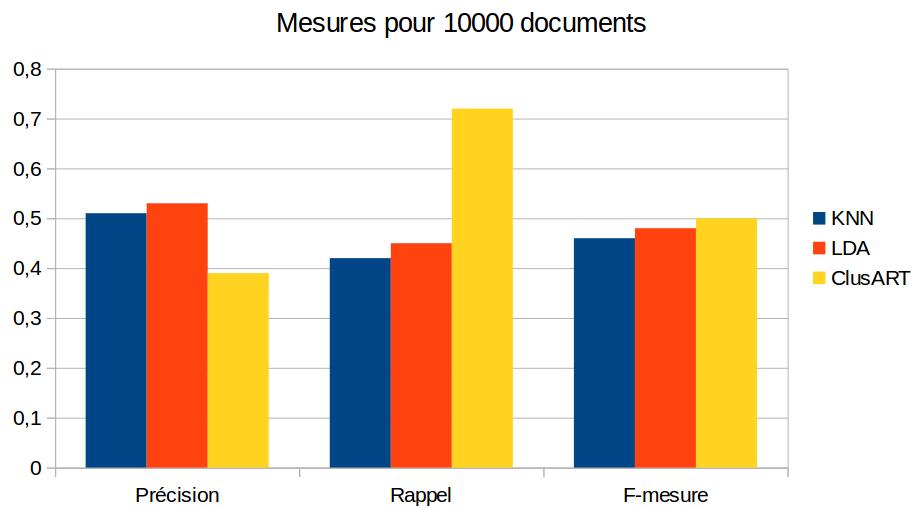}}
\captionof{figure}{Mesure de performance pour 10000 documents}
\end{figure}
\bigskip

Pour la suite des tests, nous avons augmenté le nombre de documents à 10000 documents. La figure 4.3 et le tableau 4.4 nous montrent les valeurs de rappel, précision et F-mesure pour nos 10000 documents. Les histogrammes indiquent qu'avec l'augmentation du nombre de documents ClusART surpasse toujours KNN et LDA. En fait leurs valeurs de rappel sont beaucoup plus faibles que celle obtenue par notre approche. La performance de notre approche est toujours confirmée en terme de F-mesure, elle surpasse aussi KNN et LDA. Ce que nous pouvons remarquer aussi c'est que les valeurs de rappel, précision et F-mesure ont baissés pour ClusART, KNN et LDA. 

\subsection*{- Mesures pour 15000 documents}

\begin{table}[H]
\centering
\caption{Valeurs de rappel, précision et F-mesure obtenues pour 15000 documents}
\label{my-label}
\begin{tabular}{l|l|l|l|}
\cline{2-4}
                              & Précision    & Rappel        & F-mesure      \\ \hline
\multicolumn{1}{|l|}{KNN}     & 0,48         & 0,37          & 0,41          \\ \hline
\multicolumn{1}{|l|}{LDA}     & \textbf{0,5} & 0,43          & 0,46          \\ \hline
\multicolumn{1}{|l|}{ClusART} & 0,37         & \textbf{0,69} & \textbf{0,48} \\ \hline
\end{tabular}
\end{table}

\begin{figure}[H]
\centering 
\resizebox{\linewidth}{!}{\includegraphics[]{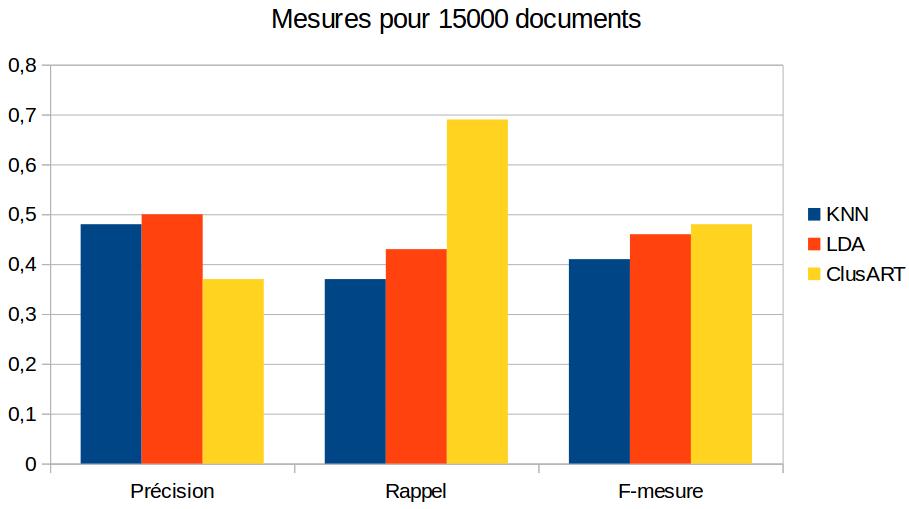}}
\captionof{figure}{Mesure de performance pour 15000 documents}
\end{figure}
\bigskip

Dans la suite de nos tests, nous avons augmenté encore le nombre de documents à 15000 documents. Selon la figure 4.4 et le tableau 4.5 nous pouvons remarquer qu'en dépit de la baisse des valeurs de précision, rappel et F-mesure; ClusART dépasse KNN et LDA. Même si KNN et LDA ont des valeurs de précision plus élevées que celle obtenue par ClusART, ce dernier les surpasse en terme de rappel. La performance de notre approche pour 15000 documents est confirmée en terme de F-mesure.

\subsection*{- Mesures pour 20000 documents}

\begin{table}[H]
\centering
\caption{Valeurs de rappel, précision et F-mesure obtenues pour 20000 documents}
\label{my-label}
\begin{tabular}{l|l|l|l|}
\cline{2-4}
                              & Précision     & Rappel        & F-mesure      \\ \hline
\multicolumn{1}{|l|}{KNN}     & 0,46          & 0,34          & 0,39          \\ \hline
\multicolumn{1}{|l|}{LDA}     & \textbf{0,47} & 0,41          & 0,43          \\ \hline
\multicolumn{1}{|l|}{ClusART} & 0,36          & \textbf{0,62} & \textbf{0,45} \\ \hline
\end{tabular}
\end{table}

\begin{figure}[H]
\centering 
\resizebox{\linewidth}{!}{\includegraphics[scale=0.8]{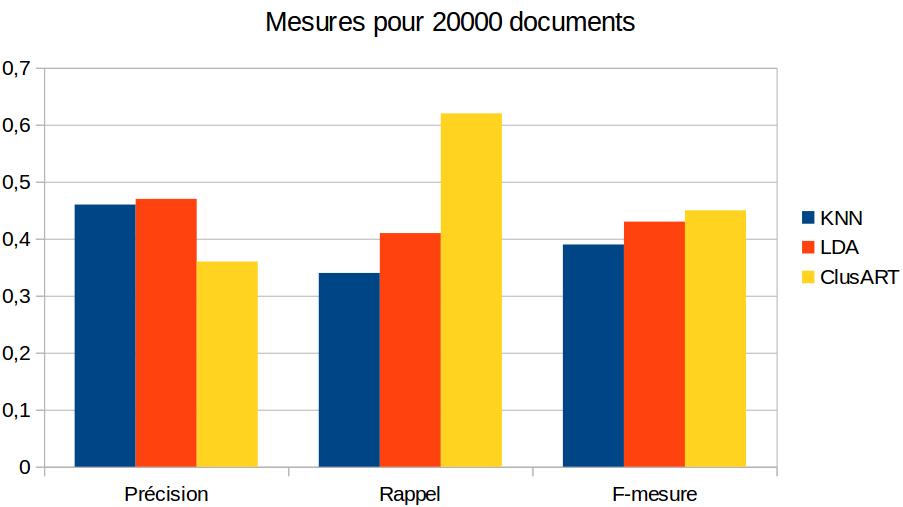}}
\captionof{figure}{Mesure de performance pour 20000 documents}
\end{figure}
\bigskip

La figure 4.5 et le tableau 4.6 décrivent les mesures de rappel, précision et F-mesure prises pour 20000 documents. Suivant les histogrammes esquissés, nous pouvons encore une fois souligner que ClusART surpasse KNN et LDA. Avec une valeur de rappel plus élevée que celles de KNN et LDA. Il est à noter que toutes les valeurs de rappel, précision et F-mesure ont encore une fois baissées suite à l'augmentation du nombre de documents. La performance de ClusART est toujours confirmée en terme de F-mesure.

\subsection*{- Récapitulation}

Pour faire un petit récapitulatif sur l'ensemble des tests effectués, nous remarquons que plus nous augmentons le nombre de documents plus les trois mesures de précision, rappel et F-mesure baissent pour KNN, LDA et ClusART.\medskip 

Pour la mesure de précision KNN et LDA dépassent ClusART. En terme de rappel, les valeurs obtenues par ClusART sont plus élevées que celles obtenues par KNN et LDA. Nous pouvons conclure que ClusART est plus performante que KNN et LDA, et cela est confirmé par les valeurs de F-mesure obtenues où ClusART surpasse à chaque fois KNN et LDA. Dans ce cas notre approche peut détecter des sujets presque pertinents.

\addcontentsline{toc}{section}{Conclusion}
\section*{Conclusion}
Dans ce chapitre, nous avons détaillé les différentes expérimentations effectuées pour la validation et l’évaluation de notre approche. L’expérimentation comparative de notre approche sur le jeu de données réelles 20 Newsgroups a montré que notre approche est capable de détecter des sujets presque pertinents. 

\newpage
\addcontentsline{toc}{chapter}{Conclusion générale}
\chapter*{Conclusion générale et perspectives}

En raison de l'énorme quantité d'informations produites à un rythme de plus en plus rapide, le web grandit tous les jours et contient d'énormes quantités de données. Ainsi, de nombreuses techniques ont été présentées pour traiter ce grand nombre de documents. Cependant, le vrai défi est de gérer ces documents en fonction de leur contenu, notamment thématique. Pour cette raison, la détection et la classification des sujets suscitent beaucoup d'intention dans les domaines de recherche traitant de différents types de documents. La détection d'un sujet est le processus d'identification du sujet d'une unité textuelle qui peut être un paragraphe, un segment ou un document texte entier.\bigskip

C’est dans ce cadre, que s’est inscrit notre travail de mémoire de mastère, où nous nous somme proposés de présenter une nouvelle approche de détection de sujets appelée ClusART. Ainsi, nous avons proposé une approche se déroulant en trois phases, à savoir: une première phase durant laquelle un pré-traitement lexical a été conduit. Une deuxième phase durant laquelle s'est effectué la construction et la génération des vecteurs représentant les documents. La troisième phase est elle même composée de deux étapes. La première étape été dédiée à l'apprentissage avec l'algorithme Fuzzy-ART. Durant la deuxième étape nous avons utilisé un classifieur utilisant Paragraph Vector pour la phase de test.\bigskip

Afin de valider et d’évaluer notre approche, nous avons mené une série d’expérimentations sur une collection de données réelles:  20 Newsgroups. Cette expérimentation a montré que notre approche est capable de détecter des sujets presque pertinents.\bigskip

Par ailleurs, notre travail peut s’étendre vers de nouvelles perspectives et il peut être encore enrichi par d’autres travaux portant principalement sur les trois aspects suivants: 
\begin{itemize}
\item La sélection des noms: il serait intéressant d'utiliser un algorithme pour filtrer les caractéristiques inutiles du texte et s'assurer que seul le nom est sélectionné. Le clustering de textes basé sur des noms peut être efficace car les noms peuvent représenter des incidents spécifiques et des événements généraux dans les textes et produire plus de bons sujets que les verbes.
\item Prendre en considération la cooccurrences des mots: les relations de cooccurrence à travers le document sont généralement négligées, ce qui conduit à la détection d'informations incomplètes.
\item Utilisation de ClusART dans un environnement multilingue.
\end{itemize}

\nocite{*}
\bibliography{biblio2}
\newpage
\input{abstract.tex}

\end{document}

%% file: abstract.tex
\thispagestyle{empty}
\paragraph*{\\ \\ Résumé \\ \\}
La détection des sujets devient importante en raison de l'augmentation du nombre d'informations disponibles sous format électronique et de la nécessité de les traiter et de les filtrer. C'est dans ce contexte que s'inscrit notre mémoire de mastère, où nous avons proposé de présenter une nouvelle approche de détection de sujets appelée ClusART. Nous avons donc proposé une approche se déroulant en trois phases, à savoir : une première phase au cours de laquelle un prétraitement lexical a été effectué. Une deuxième phase au cours de laquelle la construction et la génération des vecteurs représentant les documents a été réalisée. La troisième phase est elle-même composée de deux étapes. Dans la première étape, nous avons utilisé l'algorithme FuzzyART pour la phase de formation. Dans la deuxième étape, nous avons utilisé un classifieur utilisant Paragraph Vector pour la phase de test. L'étude comparative de notre approche sur l'ensemble de données des 20 Newsgroups a montré que notre approche est capable de détecter des sujets presque pertinents.\\

\noindent
\textbf{Mots-clés :} Détection de sujets, Regroupement de textes, FuzzyART, Classifieur Paragraph Vector.\bigskip

\bigskip

\bigskip

\vspace{5pt}\hrule\vspace{1pt}

\paragraph*{\\ \\ Abstract \\ \\ }
Topic detection becomes more important due to the increase of information electronically available and the necessity to process and filter it. In this context our master's thesis work was carried out, where we proposed to present a new approach to the detection of topics called ClusART. Thus, we proposed a three-phase approach, namely: a first phase during which lexical preprocessing was conducted. A second phase during which the construction and generation of vectors representing the documents was carried out. A third phase which is itself composed of two steps. In the first step we used the FuzzyART algorithm for the training phase. In the second step we used a classifier using Paragraph Vector for the test phase. The comparative study of our approach on the 20 Newsgroups dataset showed that our approach is able to detect almost relevant topics.\\

\noindent
\textbf{Keywords :} Topic detection, Text clustering, FuzzyART, Paragraph Vector Classifier.\bigskip

\medskip

\bigskip

\vspace{6pt}\hrule